# Hadron Scattering Lengths in Lattice QCD

M. Fukugita

*Yukawa Institute for Theoretical Physics, Kyoto University*
*Kyoto 606, Japan*
*and*
*Institute for Advanced Study, Princeton*
*NJ 08540, U. S. A.*

Y. Kuramashi and M. Okawa

*National Laboratory for High Energy Physics(KEK)*
*Tsukuba, Ibaraki 305, Japan*

H. Mino

*Faculty of Engineering, Yamanashi University*
*Kofu 404, Japan*

A. Ukawa

*Institute of Physics, University of Tsukuba*
*Tsukuba, Ibaraki 305, Japan*

Abstract

Lattice QCD calculation of $s$-wave hadron scattering lengths in the channels $\pi$-$\pi$, $\pi$-$N$, $K$-$N$, $\bar{K}$-$N$ and $N$-$N$ is carried out in the quenched approximation at $\beta = 6/g^2 = 5.7$. A variant of the method of wall source is developed for this purpose, which reduces the computer time by a factor $L^3$ on an $L^3 \times T$ lattice compared to the conventional point source method and avoids the Fierz mixing problem. A version of the method in which gauge configurations are not fixed to any gauge can be extended to calculate disconnected quark loop contributions in hadron two- and three-point functions. An analytical estimate of statistical errors for this method is worked out, and the magnitude of errors without and with gauge fixing is compared for the case of $\pi$-$\pi$ four-point functions calculated with the Kogut-Susskind quark action. For $\pi$-$\pi$ scattering both $I = 0$ and 2 scattering lengths are evaluated using the Kogut-Susskind and Wilson quark actions on a $12^3 \times 20$ lattice. For the same size of lattice, $\pi$-$N$, $K$-$N$ and $\bar{K}$-$N$ scattering lenghts are calculated with the Wilson quark action. For the $\pi$-$\pi$ and $\pi$-$N$ cases simulation results are consistent with the predictions of current algebra and PCAC within one to two standard deviations up to quite heavy quark masses corresponding to $m_\pi/m_\rho \approx 0.74$, while for the $K$-$N$ and $\bar{K}$-$N$ cases the agreement is within a factor of two. For $N$-$N$ scattering a phenomenological study with one-boson exchange potentials indicate that the deuteron becomes unbound if the quark mass is increased beyond 30–40% of the physical value. Simulations with the Wilson action on a $20^4$ lattice with heavy quarks with $m_\pi/m_\rho \approx 0.74 - 0.95$ show that the nucleon-nucleon force is attractive for both spin triplet and singlet channels, and that the scattering lengths are substantially larger compared to those for the $\pi$-$\pi$ and $\pi$-$N$ cases even for such heavy quarks. Problem of statistical errors which has to be overcome toward a more realistic calculation of hadron scattering lengths is discussed.

# 1 Introduction

Calculation of scattering lengths of hadrons presents a variety of problems of physical interest in lattice QCD approach to strong interactions. For the case of $\pi$-$\pi$ and $\pi$-$N$ scattering, the experimental $s$-wave scattering lengths are of order $0.1 - 0.4$fm which are quite small compared to the sizes of the pion and nucleon (see Table 1 for experimental values). It is well known that the small values can be understood as a result of soft pion theorems[2] that follow from approximate chiral symmetry of up and down quarks realized in a spontaneously broken mode; scattering amplitudes involving pions vanish at threshold for zero pion mass, and the $s$-wave scattering lengths predicted for the experimental pion mass are in reasonable agreement with experiment as recapitulated in the last column of Table 1. Whether a dynamical calculation directly based on the QCD Lagrangian successfully explains experiment therefore provides a useful testing ground of our theoretical understanding of chiral symmetry based on lattice QCD.

The situation is quite different for nucleon-nucleon scattering. Experimentally the scattering lengths are very large, being of the order of 10fm (see Table 1). Since chiral symmetry places no constraint on the low energy behavior, the large scattering lengths are purely dynamical phenomena. So far theoretical approach to the low energy nucleon-nucleon scattering has remained at a phenomenological level in which meson exchange models with tunable coupling constants are utilized to reproduce the low energy phase shifts[3]. Deriving the large scattering lengths from first principles of QCD represents an important challenge to lattice QCD.

Calculation of scattering lengths poses several problems from the technical point of view. One of the problems is that Euclidean hadron Green's functions amenable to Monte Carlo evaluation methods of lattice QCD are generally only indirectly connected to real time scattering amplitudes. An elegant solution to this problem is the formula[4, 5] relating scattering phase shifts, and hence also scattering lengths, to the energy of two hadron states on a finite spatial lattice which can be extracted from the exponential decay of Euclidean hadron four-point functions in time[6].

Another problem, which is computationally quite troublesome, is the calculation of hadron four-point functions itself. For hadrons in a definite momentum state, such a calculation generally requires quark propagators connecting two arbitrary space-time sites. With the conventional method of point source the necessary number of quark matrix inversions equals the space-time lattice volume, which would require a prohibitively large amount of computer time. This is a novel situation quite different from a calculation of connected three-point functions of hadrons, which can be reduced to that of connected two-point functions through source methods[7]. We note that a similar difficulty arises even for two- or three-point functions when one attempts to calculate contributions involving disconnected quark loops. Typical examples are the two-quark loop amplitude for the flavor singlet $\eta'$ propagator and flavor singlet nucleon matrix elements such as the $\pi$-$N$ $\sigma$ term.

We found that a variant of the wall source technique[8, 9] allows to overcome the problem with a modest cost of computing power[10, 11, 12, 13, 14, 15, 16]. Particularly interesting is a modified version of the original proposal[8] which does not fix gauge configurations to any gauge. This version can be applied to calculate disconnected quark loop contributions in two- and three-point functions as well as hadron four-point functions. Our chief findings obtained with the method for the $\pi$-$\pi$, $\pi$-$N$ and $N$-$N$ scattering lengths have been briefly reported in Refs. [10, 11, 12]. In this article we present the full details of the method and analyses of our calculation of hadron scattering lengths. We also report additional results for $K$-$N$ and $\bar{K}$-$N$ scattering. For applications of the method to calculate disconnected quark loop contributions in hadron two- and three-point functions, we refer to our work on the $\eta'$ meson mass[13], the $\pi$-$N$ $\sigma$ term[14, 15] and the axial vector matrix elements of the proton[15, 16].

Historically the first attempt toward a lattice calculation of hadron scattering lengths was made by Guagnelli, Marinari and Parisi[17] for the $\pi$-$\pi$ and $\pi$-$N$ cases using the Kogut-Susskind and Wilson quark actions in quenched QCD. They found a finite volume shift of the energy of two-hadron states and examined the volume dependence. However, they calculated only gluon exchange diagrams, and thus the results cannot be compared with physical scattering lengths. Gupta, Kilcup, Patel and Sharpe[18, 19] developed an analysis of the physical $\pi$-$\pi$ scattering length for the $I = 2$ channel which can be calculated with the conventional wall source method. Their studies were made in quenched QCD using the Kogut-Susskind[18] and Wilson[19] quark actions, and lattice results for scattering lengths were found to be consistent with the predictions of current algebra and PCAC for the $I = 2$ channel up to quite large quark masses.

The focus of the present work consists in an extension of the previous $I = 2$ results for the $\pi$-$\pi$ scattering length to the $I = 0$ channel and also to the $\pi$-$N$ ($I = 1/2, 3/2$), $K$-$N$ and $\bar{K}$-$N$ ($I = 0, 1$) scattering. These extensions require the modified version of the wall source method developed here. We also carry out an exploratory study of the nucleon-nucleon scattering lengths. Our simulations have been made within quenched QCD at $\beta = 5.7$ on a $12^3 \times 20$ lattice for the $\pi$-$\pi$, $\pi$-$N$, $K$-$N$ and $\bar{K}$-$N$ cases, and on a $20^4$ lattice for the $N$-$N$ case. For $\pi$-$\pi$ scattering we employed both the Kogut-Susskind and the Wilson quark action, while other scattering lengths are studied with the Wilson quark action.

This paper is organized as follows. In Sec. 2 we describe the formalism for calculation of hadron scattering lengths including the Lüscher's formula [5] and our calculational technique of modified wall sources for construction of hadron four-point functions. In Sec. 3 we summarize our data sets. The results for $\pi$-$\pi$ scattering lengths are presented in Sec. 4, where we also compare Kogut-Susskind results with and without gauge fixing, and discuss the question of infrared singularities in quenched QCD associated with the degeneracy of $\eta'$ and $\pi$. Results for $\pi$-$N$, $K$-$N$ and $\bar{K}$-$N$ scattering lengths are given in Sec. 5, while those for $N$-$N$ scattering lengths are presented in Sec. 6. In Sec. 7 we discuss the issue of statistical errors in relation to a fully realistic lattice QCD calculation of hadron scattering lengths in the future. Our conclusions are summarized in Sec. 8.

Throughout the article we employ lattice units for expressing physical quantities and suppress the lattice spacing $a$ unless necessary.

# 2 Formalism and method of measurement

## 2.1 Scattering length from two-particle energy in a finite box

The basic formula for a calculation of scattering lengths was derived by Lüscher[5] who related the $s$-wave scattering length $a_0$ between the two hadrons $h_1$ and $h_2$ to the energy shift of the two hadron state at zero relative momentum confined in a finite periodic spatial box of a size $L^3$. The formula is given by

$$E_{h_1 h_2} - (m_{h_1} + m_{h_2}) = -\frac{2\pi(m_{h_1} + m_{h_2})a_0}{m_{h_1} m_{h_2} L^3}(1 + c_1\frac{a_0}{L} + c_2(\frac{a_0}{L})^2) + O(L^{-6}) \quad (1)$$

with $c_1 = -2.837297, c_2 = 6.375183$.

The energy $E_{h_1 h_2}$ is extracted from the large time behavior of the hadron four-point function with the hadrons in the zero momentum state defined by

$$C_{h_1 h_2}(t_4, t_3, t_2, t_1) = < \sum_{\vec{x}_4} \mathcal{O}_{h_2}(\vec{x}_4, t_4) \sum_{\vec{x}_3} \mathcal{O}_{h_1}(\vec{x}_3, t_3) \sum_{\vec{x}_2} \mathcal{O}_{h_2}^\dagger(\vec{x}_2, t_2) \sum_{\vec{x}_1} \mathcal{O}_{h_1}^\dagger(\vec{x}_1, t_1) > . \quad (2)$$

In order to avoid Fierz rearrangement of quark lines to be discussed in Sec. 2.4, we choose $t_4 = t+1, t_3 = t, t_2 = 1$ and $t_1 = 0$. In this case the large time behavior reads

$$C_{h_1 h_2}(t+1, t, 1, 0) = Z_{h_1 h_2} \exp(-E_{h_1 h_2} t) + \cdots, \quad (3)$$

where dotted terms denote contributions of higher excited states.

The hadron masses $m_{h_1}$ and $m_{h_2}$ needed to obtain the energy shift $\delta E = E_{h_1 h_2} - (m_{h_1} + m_{h_2})$ are calculated in the usual manner from the two-point functions,

$$C_{h_1}(t,0) = <\sum_{\vec{x}_3} \mathcal{O}_{h_1}(\vec{x}_3,t) \sum_{\vec{x}_1} \mathcal{O}_{h_1}^\dagger(\vec{x}_1,0)> = Z_{h_1} \exp(-m_{h_1}t) + \cdots, \tag{4}$$

$$C_{h_2}(t+1,1) = <\sum_{\vec{x}_4} \mathcal{O}_{h_2}(\vec{x}_4,t+1) \sum_{\vec{x}_2} \mathcal{O}_{h_2}^\dagger(\vec{x}_2,1)> = Z_{h_2} \exp(-m_{h_2}t) + \cdots. \tag{5}$$

We extract the energy shift $\delta E$ from the ratio:

$$R(t) = \frac{C_{h_1 h_2}(t+1,t,1,0)}{C_{h_1}(t,0)C_{h_2}(t+1,1)} = \frac{Z_{h_1 h_2}}{Z_{h_1} Z_{h_2}} \exp(-\delta E t) + \cdots. \tag{6}$$

## 2.2 Hadron four-point functions

Consider $s$-wave $\pi$-$\pi$ scattering, for which isospin 0 and 2 channels are allowed due to Bose symmetry. We construct the two-pion operators for these isospin eigenchannels as

$$\mathcal{O}_{I=0}^{\pi\pi}(t) = \frac{1}{\sqrt{3}} \left( \mathcal{O}_{\pi^+}(t)\mathcal{O}_{\pi^-}(t+1) - \mathcal{O}_{\pi^0}(t)\mathcal{O}_{\pi^0}(t+1) + \mathcal{O}_{\pi^-}(t)\mathcal{O}_{\pi^+}(t+1) \right), \tag{7}$$

$$\mathcal{O}_{I=2}^{\pi\pi}(t) = \mathcal{O}_{\pi^+}(t)\mathcal{O}_{\pi^+}(t+1), \tag{8}$$

with the pion operators defined by

$$\mathcal{O}_{\pi^+}(t) = -\sum_{\vec{x}} \bar{d}(\vec{x},t)\gamma_5 u(\vec{x},t), \tag{9}$$

$$\mathcal{O}_{\pi^-}(t) = \sum_{\vec{x}} \bar{u}(\vec{x},t)\gamma_5 d(\vec{x},t), \tag{10}$$

$$\mathcal{O}_{\pi^0}(t) = \frac{1}{\sqrt{2}} \sum_{\vec{x}} \left( \bar{u}(\vec{x},t)\gamma_5 u(\vec{x},t) - \bar{d}(\vec{x},t)\gamma_5 d(\vec{x},t) \right). \tag{11}$$

In Fig. 1 we display quark line diagrams contributing to the $\pi$-$\pi$ four-point functions, denoting them as direct (D), crossed (C), rectangular (R) and vacuum (V) diagrams (meaning of circles and bars will be explained in Sec. 2.3). The ratio $R(t)$ for isospin eigenchannels can be expressed in terms of the four amplitudes in the following combination;

$$R_{I=0}^{\pi\pi}(t) = R^D(t) + \frac{1}{2}R^C(t) - 3R^R(t) + \frac{3}{2}R^V(t), \tag{12}$$

$$R_{I=2}^{\pi\pi}(t) = R^D(t) - R^C(t). \tag{13}$$

For $\pi$-$N$ scattering the $I = 1/2$ and $3/2$ operators may be taken as

$$\mathcal{O}^{\pi N}_{I=1/2}(t) = \sqrt{\frac{2}{3}}\mathcal{O}_{\pi^+}(t+1)\mathcal{O}_n(t) - \frac{1}{\sqrt{3}}\mathcal{O}_{\pi^0}(t+1)\mathcal{O}_p(t), \tag{14}$$

$$\mathcal{O}^{\pi N}_{I=3/2}(t) = \mathcal{O}_{\pi^+}(t+1)\mathcal{O}_p(t). \tag{15}$$

For the pion operators we use (9–11). Relativistic operators for the proton ($p$) and neutron ($n$) are defined with the charge conjugation matrix $C = \gamma_4\gamma_2$ taking account of the symmetries for isospin and non-relativistic spin representations,

$$\mathcal{O}_p(t) = \sum_{\vec{x}} \varepsilon_{abc}\left[\left({}^t u^a(\vec{x},t)C^{-1}\gamma_5 d^b(\vec{x},t)\right)u^c(\vec{x},t) - \left({}^t d^a(\vec{x},t)C^{-1}\gamma_5 u^b(\vec{x},t)\right)u^c(\vec{x},t)\right], \tag{16}$$

$$\mathcal{O}_n(t) = \sum_{\vec{x}} \varepsilon_{abc}\left[\left({}^t u^a(\vec{x},t)C^{-1}\gamma_5 d^b(\vec{x},t)\right)d^c(\vec{x},t) - \left({}^t d^a(\vec{x},t)C^{-1}\gamma_5 u^b(\vec{x},t)\right)d^c(\vec{x},t)\right]. \tag{17}$$

In terms of $\mathcal{O}^{\pi N}_I$ we construct the $\pi$-$N$ four-point function according to

$$C^{\pi N}_I(t) = \frac{1}{2}\text{Tr}\left[\frac{1+\gamma_4}{2} < \mathcal{O}^{\pi N}_I(t)\mathcal{O}^{\pi N\dagger}_I(0) >\right], \tag{18}$$

where trace refers to Dirac components.

Topologically four types of diagrams contribute to $\pi$-$N$ four-point functions, direct (D), crossed (C), rectangular (R) and crossed rectangular (CR), as shown in Fig. 2. In contrast to the $\pi$-$\pi$ case, these diagrams have 6, 18, 18 and 36 members having the same quark line topology but different quark contractions. All four types of diagrams contribute to both isospin eigenchannels. Weights of the members are listed in Appendix.

For $I = 0$ and $1$ operators for $K$-$N$ and $\bar{K}$-$N$ scattering we take

$$\mathcal{O}^{KN}_{I=0}(t) = \frac{1}{\sqrt{2}}\left(\mathcal{O}_{K^+}(t+1)\mathcal{O}_n(t) - \mathcal{O}_{K^0}(t+1)\mathcal{O}_p(t)\right), \tag{19}$$

$$\mathcal{O}^{KN}_{I=1}(t) = \mathcal{O}_{K^+}(t+1)\mathcal{O}_p(t), \tag{20}$$

$$\mathcal{O}^{\bar{K}N}_{I=0}(t) = \frac{1}{\sqrt{2}}\left(\mathcal{O}_{\bar{K}^0}(t+1)\mathcal{O}_n(t) - \mathcal{O}_{K^-}(t+1)\mathcal{O}_p(t)\right), \tag{21}$$

$$\mathcal{O}^{\bar{K}N}_{I=1}(t) = \mathcal{O}_{\bar{K}^0}(t+1)\mathcal{O}_p(t), \tag{22}$$

where the nucleon operators $\mathcal{O}_p$ and $\mathcal{O}_n$ are given in (16–17) and the $K$ meson operators are defined by

$$\mathcal{O}_{K^+}(t) = \sum_{\vec{x}} \bar{s}(\vec{x},t)\gamma_5 u(\vec{x},t), \tag{23}$$

$$\mathcal{O}_{K^0}(t) = \sum_{\vec{x}} \bar{s}(\vec{x},t)\gamma_5 d(\vec{x},t), \tag{24}$$

$$\mathcal{O}_{\bar{K}^0}(t) = -\sum_{\vec{x}} \bar{d}(\vec{x},t)\gamma_5 s(\vec{x},t), \tag{25}$$

$$\mathcal{O}_{K^-}(t) = \sum_{\vec{x}} \bar{u}(\vec{x},t)\gamma_5 s(\vec{x},t). \tag{26}$$

In contrast to the $\pi$-$N$ case, $K$-$N$ scattering receives contributions only from direct (D) and crossed (C) type diagrams and $\bar{K}$-$N$ scattering only from direct and rectangular (R) ones. Weights of various quark contractions to isospin eigen amplitudes are given in Appendix.

In $s$-wave $N$-$N$ scattering spin triplet $^3S_1$ and singlet $^1S_0$ states choose isospin 0 and 1 respectively due to Fermi statistics. For these eigenchannels we use the operators,

$$\mathcal{O}(^3S_1)(t) = \frac{1}{\sqrt{2}} \left( \mathcal{O}_p(t)\mathcal{O}_n(t+1) - \mathcal{O}_n(t)\mathcal{O}_p(t+1) \right), \tag{27}$$

$$\mathcal{O}(^1S_0)(t) = \mathcal{O}_p(t)\mathcal{O}_p(t+1), \tag{28}$$

where the upper two Dirac components of the nucleon operators $\mathcal{O}_p$ and $\mathcal{O}_n$ as defined in (16–17) are combined to form the respective spin states. $N$-$N$ four-point functions require only two types of diagrams shown in Fig. 3. The direct (D) diagram has 36 members with different quark contractions. For the crossed (C) diagram, which has 324 members, diagrams with a single quark-antiquark exchange and a double quark-antiquark exchange are equivalent because the two nucleon operators are summed over all spatial sites for $s$-wave states. Since quark contractions for $N$-$N$ four-point functions are very complicated and tedious, we employed symbolic manipulations on computers to work out the weights of amplitudes.

## 2.3 Wall source method without gauge fixing

Let us consider the numerical procedure for a calculation of hadron four-point functions. For $\pi$-$\pi$ scattering shown in Fig. 1 the direct and crossed diagrams can be easily calculated because we need only two quark propagators with wall sources placed at the fixed time slices $t_1$ and $t_2$[18, 19] in order to construct the corresponding four-point amplitudes for arbitrary values of $t_3$ and $t_4$ (the $N$-$N$ case is similar). This does not apply to the case of rectangular and vacuum diagrams which require additional quark propagators connecting the time slices $t_3$ and $t_4$. The same difficulty also exists in the $\pi$-$N$, $K$-$N$ and $\bar{K}$-$N$ cases for the rectangular and crossed-rectangular types of diagrams.

We handle this problem by calculating $T$ quark propagators on an $L^3 \times T$ lattice, each propagator corresponding to a wall source placed at the time slice $t = 0, \cdots, T-1$, which are defined by

$$\sum_{n''} D_{n',n''} G_t(n'') = \sum_{\vec{x}} \delta_{n',(\vec{x},t)}, \qquad 0 \le t \le T-1 \tag{29}$$

where $D$ denotes the quark matrix for the Wilson or the Kogut-Susskind quark action. The combination of $G_t(n)$ that we employ for hadron four-point functions are displayed in Figs. 1–3, where short bars represent the position of wall source and circles the sink. For example the $\pi$-$\pi$ rectangular diagram in Fig. 1(c) corresponds to

$$
\begin{aligned}
C^R&(t_4, t_3, t_2, t_1) = \\
&\sum_{\vec{x}_2, \vec{x}_3} < \mathrm{ReTr}[G_{t_1}^\dagger(\vec{x}_2, t_2) G_{t_4}(\vec{x}_2, t_2) G_{t_4}^\dagger(\vec{x}_3, t_3) G_{t_1}(\vec{x}_3, t_3)] >,
\end{aligned} \tag{30}
$$

where daggers mean conjugation by $\gamma_5$ for the Wilson quark action and by the even-odd parity $(-1)^n$ for the Kogut-Susskind quark action.

Using the relation $G_t(n'') = \sum_{\vec{x}} D_{n'',(\vec{x},t)}^{-1}$, we see that this prescription yields four-point amplitudes corresponding to non-local and non-gauge invariant hadron operators at the time slices where two wall sources or one wall and one sink are placed (*e.g.*, at the time slices $t_1$ and $t_4$ in (30)). These terms create gauge-variant noise. One way to suppress the noise is to fix gauge configurations to some gauge as is done in all recent work using wall sources[9]. A potential drawback is that gauge non-invariant states may contaminate the four-point function.

Alternatively one can choose not to fix gauge configurations to any gauge since gauge dependent fluctuations should cancel out in the ensemble average. This is the wall source version of the original proposal of extended sources[8]. One might worry if signal stands out among noise. For example, if two wall sources are placed at the same time slice, there are $O((L^3)^2)$ gauge dependent non-local terms relative to $O(L^3)$ local gauge invariant ones, and hence naively the magnitude of the noise $O(\sqrt{(L^3)^2})$ would be comparable to that of the signal. However, a generalization[20] of the well-known argument[21] can be used to show that the noise is in fact smaller than signal by a factor $L^{3/2}$ for sufficiently large $L$.

Let us illustrate this point with a simpler example of a two-point function $C_\Gamma(t)$ for the meson operator $\bar{q}\Gamma q$ for the Wilson quark action. We evaluate this quantity as

$$C_\Gamma(t) = \frac{1}{L^3} < \sum_{\vec{x}} \text{Tr} \left( G_0^\dagger(\vec{x}, t) \Gamma G_0(\vec{x}, t) \Gamma \right) >, \qquad (31)$$

where the factor $1/L^3$ is inserted to ensure a finite limit as $L \to \infty$ and the trace is taken over Dirac indeces and color indeces. Substituting the expression

$$G_0(\vec{x}, t) = \sum_{\vec{y}} G\left((\vec{x}, t), (\vec{y}, 0)\right) \qquad (32)$$

for the wall-source quark propagator $G_0(n)$ in terms of the point-to-point quark propagator $G(n, m)$, we can rewrite (31) as

$$C_\Gamma(t) = < \sum_{\vec{y}, \vec{z}} \text{Tr} \left( G_\Gamma(\vec{y}, \vec{z}) \Gamma \right) >, \qquad (33)$$

where

$$G_\Gamma(\vec{y}, \vec{z}) = \frac{1}{L^3} \sum_{\vec{x}} G\left((\vec{y}, 0), (\vec{x}, t)\right) \Gamma G\left((\vec{x}, t), (\vec{z}, 0)\right). \qquad (34)$$

The dispersion of the two-point function $C_\Gamma(t)$ is given by

$$
\begin{aligned}
\sigma^2 &= <|\sum_{\vec{y},\vec{z}} \text{Tr}\,(G_\Gamma(\vec{y},\vec{z})\Gamma)|^2> - |<\sum_{\vec{y},\vec{z}} \text{Tr}\,(G_\Gamma(\vec{y},\vec{z})\Gamma)>|^2 \\
&= <\sum_{\vec{y},\vec{z}} \text{Tr}\,(G_\Gamma(\vec{y},\vec{z})\Gamma) \sum_{\vec{z}',\vec{y}'} \text{Tr}\,(G_\Gamma(\vec{z}',\vec{y}')\Gamma)> \\
&\quad - <\sum_{\vec{y},\vec{z}} \text{Tr}\,(G_\Gamma(\vec{y},\vec{z})\Gamma)><\sum_{\vec{z}',\vec{y}'} \text{Tr}\,(G_\Gamma(\vec{z}',\vec{y}')\Gamma)>
\end{aligned}
\tag{35}
$$

where we have used the relation $G^\dagger(n,m) = \gamma_5 G(m,n)\gamma_5$ and assumed that $\gamma_5 \Gamma \gamma_5 = \pm\Gamma$. We recognize the first term on the right-hand side to be the expression for the four-point function with a quark operator placed at $\vec{z}$ and $\vec{y}'$ and an anti-quark operator at $\vec{y}$ and $\vec{z}'$ of the time slice $t = 0$. Since gauge configurations are not fixed to any gauge, non-vanishing results are obtained only if $\vec{z} = \vec{y}$ and $\vec{y}' = \vec{z}'$ or $\vec{z} = \vec{z}'$ and $\vec{y}' = \vec{y}$. We thus find that

$$
\sigma^2 = \sigma_D^2 + \sigma_C^2,
\tag{36}
$$

with

$$
\begin{aligned}
\sigma_D^2 &= \sum_{\vec{y},\vec{z}} < \text{Tr}\,(G_\Gamma(\vec{y},\vec{y})\Gamma)\,\text{Tr}\,(G_\Gamma(\vec{z},\vec{z})\Gamma)> \\
&\quad - \sum_{\vec{y},\vec{z}} < \text{Tr}\,(G_\Gamma(\vec{y},\vec{y})\Gamma)><\text{Tr}\,(G_\Gamma(\vec{z},\vec{z})\Gamma)>
\end{aligned}
\tag{37}
$$

$$
\begin{aligned}
\sigma_C^2 &= \sum_{\vec{y},\vec{z}} < \text{Tr}\,(G_\Gamma(\vec{y},\vec{z})\Gamma)\,\text{Tr}\,(G_\Gamma(\vec{z},\vec{y})\Gamma)> \\
&= \frac{1}{12} \sum_{i=S,V,T,A,P} \sum_{\vec{y},\vec{z}} < \text{Tr}\,(G_\Gamma(\vec{y},\vec{z})\Gamma\Gamma_i G_\Gamma(\vec{z},\vec{y})\Gamma\Gamma_i)>
\end{aligned}
\tag{38}
$$

where for $\sigma_C^2$ we used the Fierz transformation for Dirac and color indeces to combine the two traces into a single trace and ignored color non-singlet combinations. Substituting (34) it is easy to see that $\sigma_D^2$ equals the direct amplitude for the meson-meson scattering in the channel $\Gamma + \Gamma \rightarrow \Gamma + \Gamma$ with all meson at rest, multiplied by a factor $1/L^3$. Similarly $\sigma_C^2$ equals the crossed amplitude for the channel $\Gamma\Gamma_i + \Gamma\Gamma_i \rightarrow \Gamma + \Gamma$ multiplied by $1/L^3$. We thus find that the statistical error for the two-point function evaluated with $N_{conf}$ independent gauge configurations is

given by

$$
\begin{aligned}
\delta C_\Gamma(t) &= \sqrt{\frac{\sigma^2}{N_{conf}}} \\
&= \sqrt{\frac{1}{N_{conf}L^3}\left(C^D_{\Gamma+\Gamma\to\Gamma+\Gamma}(t) + \frac{1}{12}\sum_{i=S,V,T,A,P}C^C_{\Gamma\Gamma_i+\Gamma\Gamma_i\to\Gamma+\Gamma}(t)\right)}. \quad (39)
\end{aligned}
$$

This formula immediately shows that the statistical error decreases as $L^{-3/2}$ relative to the signal. The direct amplitude decreases as $\exp(-2m_\Gamma t)$ up to polynomials in $t$ with $m_\Gamma$ the mass of the meson in the channel $\Gamma$. For the crossed amplitude we expect the dominant term in the sum over $i$ to come from the channel $\pi + \pi \to \Gamma + \Gamma$ having the lightest energy. For a sufficiently large spatial size $L$ the crossed amplitude in this channel may be approximately calculated by a convolution of free propagation of two pions from the time slice $t = 0$ to $t'(0 < t' < t)$, a local $\pi\pi\Gamma\Gamma$ coupling at $t = t'$ and free propagation of two mesons in the channel $\Gamma$ from $t = t'$ to $t$. As an estimate of the relative error we then find

$$
\frac{\delta C_\Gamma(t)}{C_\Gamma(t)} \propto \sqrt{\frac{1}{N_{conf}L^3}}\exp((m_\Gamma - m_\pi)t). \quad (40)
$$

It is straightforward to generalize the argument to hadron four-point functions $C_{h_1 h_2}$. The relative error is given by

$$
\frac{\delta C_{h_1 h_2}(t)}{C_{h_1 h_2}(t)} \propto \sqrt{\frac{1}{N_{conf}L^3}}\exp((m_{h_1} + m_{h_2} - (n_{h_1} + n_{h_2})m_\pi)t). \quad (41)
$$

where $n_h = 1$ for mesons and $3/2$ for baryons.

We should note that the magnitude of the proportionality constant in (40) and (41) is not known, and hence the magnitude of errors may vary depending on the channel. With a practically manageable statistics of a few hundred configurations, we found that statistical errors are small for pion two- and four-point functions, whereas signal deteriorates for $\rho$ and nucleon. For nucleon, in particular, errors are significant even for the propagator. We, therefore, use the Coulomb gauge fixing for the nucleon source for calculation of scattering lengths involving nucleon.

## 2.4 Avoidance of Fierz contaminations

Our method of calculating quark propagators for all possible temporal positions of wall sources has another advantage. Suppose that we only have the quark propagator corresponding to a single wall source placed at the time slice $t$. Calculation of meson four-point functions in this case necessarily involve two meson operators $\sum_{\vec{x}, \vec{y}, a} \bar{q}(\vec{x}, t)^a \Gamma q(\vec{y}, t)^a$ and $\sum_{\vec{x}', \vec{y}', b} \bar{q}(\vec{x}', t)^b \Gamma' q(\vec{y}', t)^b$ placed at the same time slice $t$. It is easy to see that there are two contributions, one in which each of the pairs $(\bar{q}(\vec{x}, t)^a, q(\vec{y}, t)^a)$ and $(\bar{q}(\vec{x}', t)^b, q(\vec{y}', t)^b)$ form color singlets, and the other in which the Fierz-rearranged pairs $(\bar{q}(\vec{x}, t)^a, q(\vec{y}', t)^b)$ and $(\bar{q}(\vec{x}', t)^b, q(\vec{y}, t)^a)$ form color singlets. Such a rearrangement causes a mixing of diagrams having different quark line topologies. In general it also switches the spin-parity of the two hadrons in the initial state. This leads to quite a complication of analyses, especially for the Kogut-Susskind quark action, as was discussed in detail in Ref. [18]. Disentangling the two contributions is possible in principle, but difficult in practice. This, in fact, is the reason why the work of Ref. [18] could not determine the direct and crossed amplitudes separately for the $\pi$-$\pi$ four-point function for the Kogut-Susskind quark action.

The problem can be trivially solved in our method since we calculate quark propagators for all possible temporal positions of wall source, and therefore hadron operators can be placed at different time slices. In practice we separate hadron operators by a unit time slice both for the initial and final pairs of hadrons.

# 3 Data sets

In Table 2 we list the parameters of our simulation. All of our calculations are made in the quenched approximation at $\beta = 5.7$. The inverse lattice spacing determined from the $\rho$ meson mass in the chiral limit equals $a^{-1} = 1.44(2)$GeV for the Wilson quark action and $a^{-1} = 0.98(11)$GeV for the Kogut-Susskind quark action. We mostly employed a $12^3 \times 20$ lattice for $\pi$-$\pi$, $\pi$-$N$, $K$-$N$ and $\bar{K}$-$N$ scattering and a $20^4$ lattice for $N$-$N$ scattering, anticipating large scattering lengths for the latter. The lattice size must be large enough so that we can employ as weak coupling as possible to avoid finite lattice spacing effects, yet it should not be too large so as not to spoil a detection of a small energy shift of $O(L^{-3})$ predicted by the Lüscher's relation (1); our parameters are a reasonable compromise with the present computing resources. Gauge configurations are generated with a 5-hit pseudo heat bath algorithm, discarding 1000 configurations for thermalization and employing every 1000th configuration for hadron four-point function analyses.

For the $\pi$-$\pi$ four-point function we made bulk of our calculations employing wall sources without gauge fixing. A calculation with Coulomb gauge fixing was also carried out for a subset of configurations with the Kogut-Susskind quark action in order to compare the two methods (see Sec. 4.1.2). In the case of $\pi$-$N$ scattering the pion source is treated with the wall source method without gauge fixing, while for the nucleon source placed at $t = 0$ we fixed the $t = 0$ time slice of gauge configurations to Coulomb gauge in order to enhance signal to noise ratio. The $K$-$N$ and $\bar{K}$-$N$ four-point functions are calculated in the same way. We used the same quark mass for strange and up-down quarks for this case. For calculations of the $N$-$N$ four-point functions gauge configurations are fixed to Coulomb gauge over the entire space-time lattice as this leads to better signals.

Quark propagators are calculated with the Dirichlet boundary condition in the temporal direction in order to avoid contamination from hadrons propagating backward in time. The periodic boundary condition is used in the spatial directions. For the Kogut-Susskind quark action the standard conjugate gradient algorithm is employed for inverting the quark matrix, and for Wilson quark action the ILUCR method[22]. The stopping condition is chosen to ensure a 0.1% accuracy in hadron four-point functions for each configuration.

In Table 3 we summarize $\pi, \rho$ and nucleon $(N)$ masses in lattice units obtained on our ensemble of configurations through standard single exponential fits of hadron propagators over the time interval $6 \leq t \leq 12 - 14$. Also listed are the values for the pion decay constant obtained with the tadpole-improved renormalization factor[25] using $\alpha_{\overline{MS}}(1/a) = 0.2207$ for the coupling constant. The values listed are in agreement with the results obtained on larger lattices[23, 24], which are also given in Table 3. A necessary condition for the applicability of the formula (1) is that the lattice size $L$ is sufficiently large so that finite-size effects for hadron masses are negligible. The agreement provides a check, albeit indirect, on this point.

We estimate errors by the single elimination jack-knife procedure for all quantities including hadron masses, energy shifts and scattering lengths obtained by fits of two- and four-point functions.

# 4 $\pi$-$\pi$ scattering lengths

## 4.1 Kogut-Susskind quark action

### 4.1.1 No gauge fixing

Current algebra and PCAC predict[2] that the $s$-wave $\pi$-$\pi$ scattering lengths, to leading order in $m_\pi$, take the values,

$$a_0^{I=0} = +\frac{7}{32\pi}\frac{m_\pi}{f_\pi^2}, \tag{42}$$

$$a_0^{I=2} = -\frac{2}{32\pi}\frac{m_\pi}{f_\pi^2}. \tag{43}$$

It has been shown in Ref. [18] that these results can also be derived for the Kogut-Susskind quark action at a finite lattice spacing using Ward identities for $U(1)$ chiral symmetry under some continuity assumptions on the pion four-point function in external momenta. Whether numerical simulations yield results in agreement with (42–43) therefore provides a valuable check of the lattice method for hadron four-point functions.

The use of Kogut-Susskind action has a subtle problem concerning the interpretation of the four quark flavors corresponding to each Kogut-Susskind field. Following Ref. [18] we introduce a Kogut-Susskind field for each continuum flavor, regarding the four flavors of the Kogut-Susskind action as a four-fold duplication of each continuum flavor. In this view the expression (12–13) for the $\pi$-$\pi$ four-point function in isospin eigenchannels needs to be modified to

$$R^{\pi\pi}_{I=0}(t) = R^D(t) + \frac{N_f}{2}R^C(t) - 3N_f R^R(t) + \frac{3}{2}R^V(t), \qquad (44)$$

$$R^{\pi\pi}_{I=2}(t) = R^D(t) - N_f R^C(t). \qquad (45)$$

where the factors $N_f = 4$ compensate a different number of Kogut-Susskind flavor traces in the four types of diagrams. For the pion operator it is most natural to take the one in the Nambu-Goldstone channel corresponding to $U(1)$ chiral symmetry. This is the choice for which the current algebra result (42–43) can be derived.

We have carried out simulations on an $8^3 \times 20$ lattice and a $12^3 \times 20$ lattice, both at $\beta = 5.7$ and $m_q = 0.01$ in quenched QCD, employing the method of wall source without gauge fixing. The results on a $12^3 \times 20$ lattice have been briefly reported in Refs. [10, 11]. In Fig. 4 we recapitulate the individual ratios $R^X(t)$ ($X = D, C, R$ and $V$) on a $12^3 \times 20$ lattice plotted as functions of $t$. Good signals observed up to $t \approx 12$ for the rectangular amplitude and up to $t \approx 8$ for the vacuum amplitude demonstrate the practical applicability of the method of wall source without gauge fixing. For the direct and crossed amplitudes previous results[18] employed a single wall source at $t = 0$, and could not resolve the Fierz-rearranged mixing between the two amplitudes. As we discussed in Sec. 2.4 separate calculation of these diagrams are possible in our case since the two pion sources are placed at a unit lattice spacing apart in the time direction.

Physically the interesting features in Fig. 4 are (1) a very flat behavior of the direct amplitude showing that the interaction is weak in this channel, (2) an almost linear $t$ dependence of the crossed and rectangular amplitudes with a slope of similar magnitude but opposite sign, and (3) a small value of the vacuum amplitude. These characteristics are in agreement with expectations from chiral perturbation theory and the empirical OZI rule. We also note that the crossed and rectangular amplitudes have the same value at $t = 0$ as expected from the fact that their analytic expressions are identical at this value of $t$.

In Fig. 5 we plot the ratio $R_I(t)$ projected onto the $I = 0$ and 2 isospin channels for a $12^3 \times 20$ lattice. A decrease of the ratio for the $I = 2$ channel corresponds to a positive energy shift and hence to a repulsive interaction in this channel, while an increase of $R_{I=0}(t)$ means attraction for the $I = 0$ channel. A dip observed at $t \approx 2$ for the $I = 0$ channel becomes more pronounced on an $8^3 \times 20$ lattice as shown in Fig. 6(note the change of vertical scale from Fig. 5). The physical origin of the dip and its size dependence is not clear to us.

Extraction of the energy shift $\delta E_I$ from $R_I(t)$ and thence the $s$-wave scattering length require some care. Since our calculations are made in quenched QCD, rescattering effects that require sea quark loops are not properly taken into account. This effect starts with terms of $O(t^2)$ in $R_I(t)$. For the Kogut-Susskind quark action there are further complications arising from the non-degeneracy of pions in the Nambu-Goldstone and other channels at a finite lattice spacing. Briefly stating, the contribution of non-Nambu-Goldstone pions in the intermediate states is exponentially suppressed for large times due to their heavier masses compared to that of the Nambu-Golstone pion, instead of yielding terms growing with powers of $t$ for the degenerate case. This affects both the relation between the ratio $R_I(t)$ and the energy shift $\delta E_I$ and that between $\delta E_I$ and scattering lengths (1). A detailed analysis[18] shows that the $O(L^{-5})$ terms in the relation (1) are invalidated by the effect, and that $O(t^2)$ terms in $R_I(t)$ are not correct even in full QCD. Thus, a proper extraction of scattering lengths for the present case requires the spatial lattice size to be large enough so that the $O(L^{-5})$ terms are small; $\delta E_I$ should also be small so that there is a range of $t$ over which $R_I(t)$ exhibits a linear behavior.

Based on the considerations above we fit $R_I(t)$ with a linear form $Z_I(1 - \delta E_I t)$ with the fitting range chosen to be $4 \leq t \leq 9$, ignoring higher order terms. The fitted values of $\delta E_I$ and the results for the $s$-wave scattering length $a_0$ in lattice units obtained using (1) are summarized in Table 4. Here we used pion mass given in Table 3. The errors quoted for $a_0$ are statistical only. For a $12^3 \times 20$ lattice the $O(L^{-5})$ term contributes 10% in the $I = 0$ channel, although it is negligibly small ($< 1\%$) for $I = 2$. The $O(t^2)$ terms in $R_I(t)$, neglected in the above procedure, contribute a few percent in both isospin eigenchannels. These uncertainties are within the statistical errors of 16% for $I = 0$ and 9% for $I = 2$.

On an $8^3 \times 20$ lattice, the $Z$ factor for the $I = 0$ channel deviate severely from unity due to a dip of $R_{I=0}(t)$ for small $t$, rendering the value of the extracted scattering length questionable. For the $I = 2$ channel, the value for $a_0$ is consistent with that from a $12^3 \times 20$ lattice.

In Table 4 we also list in brackets the values predicted by current algebra and PCAC (42–43) substituting the value of $m_\pi$ and the pion decay constant $f_\pi$ in Table 3. We observe that the simulation results are consistent with (42–43) within $1 - 1.5$ standard deviations, which we find quite reasonable in view of the systematic uncertainties discussed above. Results for $a_0(I = 2)$ in agreement with (43) was previously obtained [18] with the assumption that the contribution of the direct diagram is negligible.

### 4.1.2 Coulomb gauge fixing

We have repeated the calculation of the $\pi$-$\pi$ four-point function, fixing gauge configurations to the Coulomb gauge for all time slices for a $12^3 \times 20$ lattice. In this case the wall source and sink for pion propagators in the denominator of $R^X(t)$ have to be chosen in the same combination as for the four-point amplitude in the numerator for each type of diagrams $X = D, C, R$ and $V$. Without this matching the $Z$ factors for the ratio $R^X(t)$ would be different between the diagrams, and the combination (44–45) would no longer project out isospin eigenchannels. This problem is absent for wall sources without gauge fixing because non-local terms in the wall source operator cancel out after averaging over gauge configurations.

In Fig. 7 we compare the ratio $R_I(t)$ calculated in Coulomb gauge (open symbols) and without gauge fixing (filled symbols) for the same set of 60 configurations on a $12^3 \times 20$ lattice. We suspect that a wiggle seen up to $t \approx 4$ for the $I = 0$ channel in the Coulomb gauge calculation is due to a contribution of non-local $\rho$ that can be emitted from the gauge fixed wall source. A slight discrepancy in the intercept for the $I = 2$ channel is ascribed to the difference of $Z$ factors between the two calculations. Except for these points the two data sets are consistent. Results for scattering lengths extracted with the same procedure as for the case without gauge fixing agree with those for the case of no gauge fixing within one standard deviation (see Table 4).

An interesting question is to what extent gauge non-invariant errors are controlled in the method of wall source without gauge fixing compared to the gauge fixed case. To examine this point we plot in Fig. 8 the single-elimination jackknife errors of $R_I(t)$ for the two calculation on the same set of 60 configurations on a $12^3 \times 20$ lattice. Although the magnitude of errors is larger for the non-gauge fixed case (filled symbols), the amount of increase of errors is contained at the level of a factor of $1.5 - 2$ times those for the Coulomb gauge fixing (open symbols), showing that gauge variant noise does not give rise to a serious problem for calculation of $\pi$-$\pi$ four-point functions.

## 4.2 Wilson quark action

We have also applied the wall source method without gauge fixing to the Wilson quark action at the hopping parameter $K = 0.164$, employing 70 configurations on a $12^3 \times 20$ lattice.

In Fig. 9 we show the ratios $R^X(t)$ for $X = D, C, R$ and $V$. The direct and crossed amplitudes were previously calculated in Ref. [19] through gauge fixed wall sources placed at the same time slice. We observe in Fig. 9 that the slope for the rectangular amplitude $R^R(t)$ is quite small compared to that for the crossed amplitude $R^C(t)$. This is different from the Kogut-Susskind case for which both amplitudes exhibit a similar slope (see Fig. 4). We consider that this is due to the heavy quark mass ($m_\pi/m_\rho = 0.74$) for the present calculation with the Wilson quark action compared to the small value ($m_\pi/m_\rho = 0.33$) for the Kogut-Susskind case. Another difference between the Wilson and Kogut-Susskind results is a positive curvature for the direct amplitude seen in Fig. 9 for the Wilson case, whose origin is not clear to us. Also the errors for the vacuum diagram blows up much more rapidly (compare Fig. 9 with Fig. 4). This, however, may be ascribed to a heavier pion mass for the Wilson simulation as will be discussed in Sec. 7 below.

The ratios $R_I(t)$ for the isospin eigenchannels $I = 0$ and 2 are plotted in Fig. 10. To extract the energy shift $\delta E_I$ for each channel, we again employ a linear form $Z_I(1 - \delta E_I t)$ with the fitting range chosen to be $4 \leq t \leq 9$. The values of the scattering lengths are listed in Table 5, together with those predicted by current algebra (42–43) but evaluated with the measured values of $m_\pi$ and $f_\pi$ in Table 3. We quoted only the statistical errors for the scattering lengths $a_0$. Simulation results are consistent with current algebra and PCAC within 1-2 standard deviations. For the $I = 2$ channel a similar agreement was previously found in Ref. [19].

## 4.3 Question of quenched chiral divergences

All of our results are obtained in quenched QCD. A problem with quenched calculations is the uncertainty in $O(t^2)$ term of the ratio $R_I(t)$ due to lack of dynamical quark loops. We have bypassed this problem by looking only at the region of $t$ where the ratio $R_I(t)$ shows a linear behavior. Also our data do not extend to the region of large $t$, where a curvature due to $O(t^2)$ terms can be established, because of increase of errors.

Another potential problem is that the $\pi$-$\pi$ four-point function in quenched QCD could be affected by infrared divergences appearing in the chiral limit due to $\eta'$ loops[26]. These divergences originate from the double pole $m_0^2/(p^2 + m_\pi^2)^2$ in the disconnected two quark loop amplitude of the $\eta'$ propagator, where $m_0^2 = m_{\eta'}^2 - m_\pi^2$ represents the mass-squared splitting between the $\eta'$ and pseudoscalar octet mesons in the full theory. For the $\pi$-$\pi$ scattering amplitude the one-loop diagram in chiral perturbation theory formed by two double-pole propagators yields a divergent imaginary part at threshold in the $s$ channel[27]. This implies that the vacuum amplitude $R^V(t)$ would be ill-behaved as $t$ increases. A rapid loss of signal seen in Fig. 4 and Fig. 9 may be partly related to this point.

The same diagram viewed in the $t$ channel gives a contribution of the form,

$$\delta a_0 = \frac{1}{1536\pi^3} \frac{1}{m_\pi f_\pi^4} \left( \frac{m_0^2}{N_f} \right)^2 \tag{46}$$

to the $s$-wave scattering length, which diverges as $m_\pi \to 0$. In terms of the energy shift in the direct amplitude $R^D(t)$, this contribution translates to

$$\delta E_D = \frac{1}{384\pi^2} \frac{1}{m_\pi^2 f_\pi^4} \left( \frac{m_0^2}{N_f} \right)^2 \frac{1}{L^3} \tag{47}$$

with $L$ the spatial size.

Recently we have calculated[13] the parameter $m_0$ for quenched QCD with Wilson quark action using the same set of configurations on a $12^3 \times 20$ lattice as are employed in the present work. At the hopping parameter $K = 0.164$ where our scattering length calculation is made, we obtained $m_0/\sqrt{N_f} = 0.147(8)$. Combining this result with those of $m_\pi$ and $f_\pi$ in Table 3 we find $\delta E_D = 2.7 \times 10^{-6}$ for the expected magnitude of energy shift for the direct amplitude for the Wilson case, which is far too small to be detectable in our data for the ratio $R^D(t)$.

The singular contribution (47) diverges in the chiral limit. In order to see at what quark mass the contribution seriously begins to affect the chiral behavior we recall[13] that the values for $m_0$ we obtained can be fitted very well with a linear function of the quark mass $m_q = (1/K - 1/K_c)/2$;

$$\frac{m_0}{\sqrt{N_f}} = 0.299(14) - 1.57(19)m_q. \tag{48}$$

For $m_\pi^2$, $m_\rho$ and $f_\pi$ we find from the values reported in Ref. [24] that

$$m_\pi^2 = 2.72(10)m_q, \tag{49}$$

$$m_\rho = 0.5361(92) + 1.61(14)m_q, \tag{50}$$

$$f_\pi = 0.0680(28) + 0.338(43)m_q. \tag{51}$$

Combining (46–51) one can estimate the fractional change of the $I = 0$ scattering length relative to the current algebra value $a_0 = 7m_\pi/(32\pi f_\pi^2)$ expected for the Wilson action at $\beta = 5.7$. The result is plotted by a solid line in Fig. 11 as a function of $m_\pi/m_\rho$. We see that the contribution of the singular term would become significant only for realistically small values of quark masses, which are beyond our computing resources.

A similar estimate for the Kogut-Susskind case, strictly speaking, requires results for $m_0$ for that action which is not available. One may, however, take over (48) as a guide. Employing the data of Ref. [28] for the Kogut-Susskind quark action at $\beta = 5.7$, which yield,

$$m_\pi^2 = 0.004(1) + 7.96(5)m_q, \tag{52}$$

$$m_\rho = 0.782(86) + 10.0(6.7)m_q, \tag{53}$$

$$f_\pi = 0.118(7) + 0.97m_q, \tag{54}$$

one finds the dotted line in Fig. 11 for the fractional change $\delta a_0/a_0$ for the Kogut-Susskind case. Again the singular contribution is very small $\delta a_0/a_0 \approx 0.001$ at the point $m_q = 0.01$ $(m_\pi/m_\rho = 0.33)$ of the simulation. For the energy shift we find $\delta E = 3.8 \times 10^{-5}$, which may be compared with the value $\delta E = 1.2(4.0) \times 10^{-4}$ extracted from the slope of the direct amplitude $R^D(t)$.

The estimate above indicates that the spurious infrared divergence and possible failure of the quenched approximation do not become manifest in $\pi$-$\pi$ four-point functions unless the simulation is made with a quark mass much smaller than the value being taken in current quenched QCD studies. Of course, the problem of unphysical singularity does not arise if calculations are made on full QCD gauge configurations.

## 4.4 Summary of $\pi$-$\pi$ scattering lengths

Our results for $\pi$-$\pi$ scattering lengths are summarized in Fig. 12 in terms of the dimensionless ratio $32\pi f_\pi^2 a_0/m_\pi$ together with those of Ref. [18, 19] for $I = 2$. We observe an agreement of lattice results with current algebra predictions up to quite heavy quark masses $(m_\pi/m_\rho \approx 0.7 - 0.8)$ for both $I = 0$ and 2 channels. In detail, however, the Wilson result for the $I = 0$ scattering length obtained at a heavy quark mass is somewhat smaller than the current algebra value, while the Kogut-Susskind results at a small quark mass are larger. It would be an interesting problem to repeat the Kogut-Susskind simulation for larger quark masses in order to see if the trend seen for the Wilson action is reproduced. For the $I = 2$ channel we do not observe such a difference between the two actions.

Aside from the complications for the Kogut-Susskind case due to breaking of flavor symmetry discussed in Sec. 4.1.1, possible sources of systematic errors in our results are the uncertainties in higher order $1/L$ terms of (1) and scaling violations due to a fairly large lattice spacing of our simulation ($a^{-1} \approx 1-1.5$ GeV at $\beta = 5.7$ determined from the $\rho$ meson mass[23, 24]). For the $I = 2$ channel the previous work[18, 19] checked these points. In particular, a comparison of results at $\beta = 5.7$ and 6.0 indicates that finite lattice spacing errors are not large, at least for the Kogut-Susskind case, for the $I = 2$ channel (compare open squares ($\beta = 5.7$) and open diamonds ($\beta = 6.0$) in Fig. 12). Whether this applies also to the $I = 0$ channel should be examined in future work.

## 5 $\pi$-$N$, $K$-$N$ and $\bar{K}$-$N$ scattering lengths

### 5.1 $\pi$-$N$ scattering lengths

We calculate the $\pi$-$N$ scattering lengths from $\pi$-$N$ four-point functions. Since the wall source technique without gauge fixing does not yield good signals for the nucleon, we use the Coulomb gauge fixing at the $t = 0$ time slice for the nucleon source. We employ the Wilson quark action at hopping parameters $K = 0.164$ and $K = 0.1665$, using 60 and 30 gauge configurations, respectively.

In Fig. 13 we show the ratio $R_I(t)$ of $\pi$-$N$ amplitude for the isospin eigenchannels $I = 1/2$ and $3/2$ at the two values of the hopping parameter. The data are fitted with the linear form $R_I(t) = Z_I(1 - \delta E_I t)$ over $4 \leq t \leq 9$, and the results for the scattering length are listed in Table 6. They are compared with predictions of current algebra and PCAC (values in brackets in Table 6), which are given by

$$a_0(I = 1/2) = +\frac{1}{4\pi}\frac{\mu_{\pi N}}{f_\pi^2}, \tag{55}$$

$$a_0(I = 3/2) = -\frac{1}{8\pi}\frac{\mu_{\pi N}}{f_\pi^2} \tag{56}$$

where $\mu_{\pi N} = m_\pi m_N/(m_\pi + m_N)$ is the reduced mass and the right-hand sides are estimated with the measured values of $m_\pi$, $m_N$ and $f_\pi$ in Table 3. The results at $K = 0.164$ are consistent with the current algebra prediction, in spite of the fact that simulations are made with quite heavy quarks corresponding to $m_\pi/m_\rho = 0.74$. The quality of data deteriorates for a lighter quark of $K = 0.1665$; more statistics are clearly needed as Fig. 13 indicates.

Let us define $R_\pm^X(t)$ to be the contribution of $X$ type diagrams ($X = D, C, R$ and $CR$; see Fig. 2) to crossing even ($+$) and odd ($-$) channels, summed over fermion contractions having the quark line topology of the $X$ diagram. In terms of the isospin eigen amplitudes $R_I(t)$, the crossing even and odd amplitudes are given by

$$R_+(t) = \frac{1}{3}R_{I=1/2}(t) + \frac{2}{3}R_{I=3/2}(t), \tag{57}$$

$$R_-(t) = \frac{1}{3}R_{I=1/2}(t) - \frac{1}{3}R_{I=3/2}(t). \tag{58}$$

The ratios $R_\pm^X(t)$ are plotted in Fig. 14 as a function of $t$ for $K = 0.164$. We recall that the crossed ($C$) and rectangular ($R$) amplitudes are governed by current algebra; the soft pion theorem predicts that the crossing even amplitudes $R_+^R(t)$ and $R_+^C(t)$ should show a reversal of sign in the slope with respect to $t$. The direct ($D$) and crossed rectangular ($CR$) amplitudes do not appear in current algebra to leading order in $m_\pi$, and therefore we expect that $R_\pm^D(t)$ and $R_\pm^{CR}(t)$ are flat. Our results in Fig. 14 are roughly consistent with these expectations except that the $t$ dependence is not very pronounced for the rectangular diagram compared to the crossed diagram and that the direct amplitude exhibits a positive curvature. Both these features are similar to what appeared in the $\pi$-$\pi$ case with the Wilson quark action, perhaps originating from heavy quark mass employed ($m_\pi/m_\rho = 0.74$).

## 5.2  $K$-$N$ and $\bar{K}$-$N$ scattering lengths

The prediction of current algebra and PCAC for SU(3) symmetry is given by

$$a_0^{KN}(I=0) = 0, \tag{59}$$

$$a_0^{KN}(I=1) = -\frac{1}{4\pi}\frac{\mu_{KN}}{f_K^2}, \tag{60}$$

$$a_0^{\bar{K}N}(I=0) = \frac{3}{8\pi}\frac{\mu_{KN}}{f_K^2}, \tag{61}$$

$$a_0^{\bar{K}N}(I=1) = \frac{1}{8\pi}\frac{\mu_{KN}}{f_K^2} \tag{62}$$

with $\mu_{KN} = m_K m_N / (m_K + m_N)$ being the reduced mass. Looking at Table 1 we observe that the results for the $KN$ channel are not unreasonable compared with experiment, although the $I = 1$ scattering length is off by a factor of two probably due to a long extrapolation to the soft $K$ meson limit. On the other hand, comparison of the predictions (61–62) for the $\bar{K}$-$N$ channel with the experiment requires a caution: $\pi\Lambda$ and $\pi\Sigma$ channels are already open at the $\bar{K}N$ threshold. It is interesting to note that the prediction (62) for the $I = 1$ $\bar{K}N$ scattering length still agrees with the real part of the experimental value. The fact that the real part of the $I = 0$ $\bar{K}N$ scattering length predicted by (61) has a sign opposite to that of the experiment is clearly due to the presence of $\Lambda(1405)$. We point out that the lattice results reported below can be compared better with the current algebra values, since the present lattice calculation with the quenched approximation is basically a single channel calculation without taking into account an opening of the channels below the threshold.

Our results for $K$-$N$ and $\bar{K}$-$N$ scattering lengths are obtained at $K = 0.164$, employing the same hopping parameter for the strange and up-down quark masses. The contribution of the direct and crossed diagrams to the $K$-$N$ amplitude in the $I = 0$ and 1 isospin eigenchannels is plotted in Fig. 15. For the $I = 0$ channel (filled circles) a small positive slope of the direct amplitude is almost canceled out by the negative slope of the crossed one. The direct and rectangular contribution to the $\bar{K}$-$N$ amplitude for both isospin eigenchannels are presented in Fig. 16.

The ratio $R_I(t)$ for the isospin eigenchannels is shown for the $K$-$N$ and $\bar{K}$-$N$ cases in Fig. 17 and 18 respectively. We made a linear fit $R_I(t) = Z_I(1 - \delta E_I t)$ over $4 \leq t \leq 9$ with the results given in Table 7, together with the current algebra values evaluated with the measured value of $m_\pi, m_N$ and $f_\pi$ in Table 3. We see that all signs agree with those of the current algebra results. For the $I = 0$ $KN$ scattering length for which the current algebra value is vanishing, the lattice result also yields a smaller value. For other cases, the agreement is up to a factor of two.

# 6    $N$-$N$ scattering lengths

Calculation of $N$-$N$ scattering lengths poses a number of challenging obstacles not encountered for the case of $\pi$-$\pi$ and $\pi$-$N$ scattering. First of all the experimental scattering lengths are quite large, being of order 10fm (see Table 1). This means that lattice sizes much larger than $2 - 3$fm, which are accessible in current numerical simulations, will be needed to suppress $O(L^{-6})$ corrections neglected in the Lüscher's formula (1). Secondly extraction of the scattering length in the spin triplet channel requires a calculation of the lowest scattering state orthogonal to the ground state, which is the deuteron bound state. In fact the negative sign of the $s$-wave spin triplet scattering length is a consequence of the existence of the deuteron as follows from Levinson's theorem. Finally statistical fluctuations in $N$-$N$ four-point functions are expected to grow more rapidly than in the $\pi$-$\pi$ and $\pi$-$N$ cases toward large time separations and small quark masses as we shall discuss in detail in Sec. 7 below.

A possible strategy in this situation is to start from the region of heavy quark masses, where simulations are easier, and subsequently reduce the quark mass. To follow this approach, we first examine the behavior of nucleon-nucleon forces toward larger quark masses through a phenomenological model of one-boson exchange potentials. We then present our lattice results carried out with heavy quarks corresponding to $m_\pi/m_\rho \approx 0.74 - 0.95$.

## 6.1    Phenomenological considerations

Low energy $N$-$N$ scattering data are usually described by phenomenological models of one-boson exchange potentials[3]. At large nucleon-nucleon separations $r \gtrsim 2$fm, the potential is dominated by one-pion exchange. At intermediate dis-tances $2 \gtrsim r \gtrsim 1$fm, the two-pion exchange provides the dominant attraction, which is modelled by an isoscalar scalar particle $\sigma$ of mass $\approx 500$MeV. Heavier bosons such as $\rho$ and $\omega$ also become important in this range of distance. The potential turns into a hard core at $r \approx 0.5$fm whose dynamical details are not understood well.

The formation of bound states in the triplet $^3S_1$ channel that emerges from such a potential is subtle. The central potential, which has a depth of about $-50$MeV at $r \approx 1$fm both in the triplet $^3S_1$ and singlet $^1S_0$ channels, is insufficient for bound state formation. For the triplet channel, however, the tensor potential provides an additional attraction of a similar depth. This leads to the binding of the deuteron with a small binding energy of 2.22457MeV with a sizable mixing of $D$ wave.

Let us consider what happens if the quark mass is increased from the physical value. The mesons most affected by the increase is the pion and $\sigma$ representing the two-pion exchange, whose masses increase as $\sqrt{m_q}$. Thus the attractive potential provided by these states becomes rapidly shorter ranged (while the hard core mostly given by $\omega$ changes little), and the two nucleons have to come closer to remain in the potential well. One can imagine that the resulting increase of kinetic energy may easily overwhelm the small deuteron binding energy, leading to unbinding of the deuteron state.

We have examined whether this actually takes place by employing the one-boson exchange model of Ref. [29, 30] and varying the values of $m_\pi$ and $m_\sigma$ according to our lattice result for the slope $m_\pi^2/m_q$. We find a divergence of the scattering length for the triplet channel taking place at $m_q = 6.3$MeV that signals unbinding of the deuteron. The fractional increase of quark mass from the value 4.9MeV corresponding to the physical point is only 30%. We have repeated the analysis in several alternative ways, *e.g.,* varying other meson masses and/or the nucleon mass assuming the relation $m_H = A_H + B_H m_q$, and also including a variation of the meson-nucleon coupling constants. We have found that the qualitative features are not modified: the deuteron most probably ceases to exist as a bound state when the quark mass is increased by more than 30–40% from the physical value. Another common feature is that the scattering lengths decrease to a value of the order of 1fm, when quark mass is increased so that $m_\pi/m_\rho \approx 0.7 - 0.9$. It is reasonable to expect that the scattering length takes a value similar to the hadron size in the absence of bound-state effects.

## 6.2  Lattice results

Our lattice study is carried out with the Wilson quark action. In the calculation of the nucleon four-point function gauge configurations are fixed to the Coulomb gauge over all space-time to enhance signals. Projecting out spin singlet and triplet combinations of the two nucleon system is made by non-relativistically combining the upper Dirac components of the nucleon operator as explained in Sec. 2.2. Other technical features are similar to the $\pi$-$\pi$ and $\pi$-$N$ cases.

In Fig. 19 we show the individual contributions of the direct and crossed amplitudes to the ratio $R(t)$ for the triplet $^3S_1$ (filled circles) and singlet $^1S_0$ (open triangles) channels obtained at $K = 0.160$. We find that the direct amplitudes for the two channels are virtually the same. On the other hand, the crossed amplitude in the spin triplet channel increases in contrast to a flat behavior in the spin singlet channel.

In Fig. 20 we show $R(t)$ for the spin singlet and triplet channels at $K = 0.160$, which corresponds to $m_\pi/m_\rho = 0.85$. A clear signal with a positive slope is observed for both channels, which means attraction ($\delta E_{NN} = E_{NN} - 2m_N < 0$). Similar results are obtained at two other values of the hopping parameter $K = 0.150(m_\pi/m_\rho = 0.95)$ and $0.164(m_\pi/m_\rho = 0.74)$. We extract the energy shift $\delta E_{NN}$ by fitting $R(t)$ to a linear form $R(t) = Z(1 - \delta E_{NN}t)$. The fitting range is chosen to be $4 \leq t \leq 9$ for $K = 0.150$ and $0.160$. For the case of $K = 0.164$, however, we used $2 \leq t \leq 6$ due to poor quality of our data. The fitted values of $\delta E_{NN}$ are quite small ($\approx 0.01$), justifying the use of a linear function instead of an exponential.

From phenomenological considerations in Sec. 6.1 we expect that the deuteron is not a bound state at a heavy quark mass where our simulations are made. We then extract the scattering lengths through (1) for both spin singlet and triplet channels. The results in lattice units are tabulated in Table 8. We should remark that the scattering lengths we found are large enough to warrant a calculation with yet a larger lattice size, even though we used quite a large size of $20^4$: the three terms in (1) are comparable in magnitude. A finite-size analysis with larger lattice sizes is clearly necessary.

In Fig. 21 we compare the results for the $N$-$N$ scattering lengths with those for $\pi$-$\pi$ and $\pi$-$N$ obtained with the Wilson quark action. Conversion to physical units is made using $a = 0.137(2)$fm determined from the $\rho$ meson mass. It is apparent that the $N$-$N$ scattering lengths are substantially larger than the $\pi$-$N$ and $\pi$-$\pi$ scattering lengths already for a heavy quark corresponding to $m_\pi/m_\rho \approx 0.74$. Also noteworthy is the trend, albeit with sizable errors, that the values for the spin triplet channel are larger than those for the singlet channel, indicating a stronger attraction in the triplet channel. This is consistent with the existence of the deuteron bound state for physical quark mass.

In physical units our results correspond to $a_0(NN) \approx 1.0 - 1.5$fm. These values are small compared to the experimental value of order 10fm. However, the large experimental scattering lengths are a reflection of the fact that the $s$-wave nucleon-nucleon system is either marginally bound (triplet channel) or very close to a bound-state formation (singlet channel), which would not be the case for heavy quarks studied in our simulation. Therefore, our results are not unreasonable. We expect that lattice results will exhibit an increase as the quark mass is decreased. For the triplet channel, in particular, the scattering length should diverge at the point where the deuteron bound state is formed. According to the phenomenological consideration in the previous section, however, this will take place quite close to physical quark mass, which is not accessible in current lattice QCD simulations.

# 7    Problem of statistical errors

Our results for $R(t)$ for various channels share the feature that statistical errors grow rapidly for large times, examples of which are seen in Fig. 5 and Fig. 10 for $\pi$-$\pi$, Fig. 13 for $\pi$-$N$ and Fig. 20 for $N$-$N$ cases. This behavior can be understood from (41), which implies that the error $\delta R^X(t)$ of the ratio $R^X(t)$ for a diagram $X$ increases as $\delta R^X(t) \propto \exp(\alpha_X t)$ with the exponent $\alpha_X$ given in Table 9. The $t$ dependence of the measured errors agree very well with this formula. Some typical examples are shown in Fig. 22.

This consideration clarifies why the vacuum amplitude for the $\pi$-$\pi$ scattering rapidly loses signal as $t$ increases: the error has the largest exponent $\alpha_V = 2m_\pi$ among the four amplitudes contributing to the $\pi$-$\pi$ four-point function. We also expect that the rate of growth of errors decreases for the $\pi$-$\pi$ case toward small quark mass, since the exponent is governed by the pion mass. Exploring the light quark mass region is more difficult for $\pi$-$N$ and $N$-$N$ scattering: the exponent for these cases is a difference of the nucleon mass and pion mass with some coefficient, which depends on the channel, and the exponent stays non-vanishing or even increases toward the chiral limit. We illustrate this point in Fig. 23 where we plot the largest exponent, $2m_\pi$ for $\pi$-$\pi$, $m_N - m_\pi/2$ for $\pi$-$N$ and $2m_N - 3m_\pi$ for $N$-$N$ cases, as a function of $m_\pi/m_\rho$ where we used the fits (49–50) for $m_\pi$ and $m_\rho$ and

$$m_N = 0.741(21) + 3.80(31)m_q \tag{63}$$

for $m_N$ obtained from the hadron mass data in Ref. [24]. We observe that the problem will be particularly severe for the $N$-$N$ four-point function for which the exponent begins to increase for $m_\pi/m_\rho \lesssim 0.5$.

# 8 Conclusions

In this work we have calculated hadron scattering lengths in a variety of channels at $\beta = 5.7$ in the quenched approximation. We have demonstrated that Lüscher's formula that correlates scattering lengths with the energy of the two-hadron system confined in a finite box can be used to calculate any class of diagrams of scattering amplitudes, when implemented with a modified wall source method proposed by the present authors. The results are physically sensible and encouraging, being in a reasonable agreement with current algebra values and the experiment for $\pi$-$\pi$ and $\pi$-$N$ cases; for $N$-$N$ scattering the resulting scattering lengths are significantly larger compared to the first two cases even for heavy quarks.

In the present study dynamical effects of sea quarks are not taken into account. However, it is straightforward to apply our method once configurations are generated with dynamical quarks. Worrisome toward a more realistic calculation is the exponential growth of errors for large times discussed in Sec. 7. The problem is particularly acute for $N$-$N$ scattering, for which the rate is numerically large and is expected to increase as the chiral limit is approached. The anticipated large $N$-$N$ scattering lengths, which henceforth requires a large lattice size, makes the problem further difficult. Advance in our understanding of the low energy nucleon-nucleon scattering from the first principle has to await not only future progress of computing power but also further innovation of calculational techniques.

# Acknowledgements

Numerical calculations for the present work have been carried out on HITAC S820/80 at KEK. Three of us (MF, HM and AU) would like to thank KEK for warm hospitality. Useful discussions with C. Bernard, N. Cabbibo, M. Golterman and S. Sharpe are gratefully acknowledged. This work is supported in part by the Grants-in-Aid of the Ministry of Education (Nos. 03640270, 05640325, 05640363, 06640372, 05-7511, 05NP0601, 06NP0601).

# Appendix

In this Appendix we enumerate weights of various quark contractions contributing to the $\pi$-$N$, $K$-$N$ and $\bar{K}$-$N$ four-point functions. Let us consider the meson-baryon operators given by

$$\mathcal{O}(x, y) = \varepsilon_{abc} \left( {}^t q_{\tilde{1}}^a C^{-1} \gamma_5 q_{\tilde{2}}^b \right) q_{\tilde{3}}^c(x) \times \bar{q}_5 \gamma_5 q_{\tilde{4}}(y), \tag{64}$$

$$\mathcal{O}^\dagger(x', y') = \varepsilon_{a'b'c'} \bar{q}_3^{c'} \left( \bar{q}_2^{b'} \{C^{-1} \gamma_5\}^{\dagger t} \bar{q}_1^{a'} \right)(x') \times -\bar{q}_4 \gamma_5 q_{\tilde{5}}(y'). \tag{65}$$

where subscripts with tilde are assigned to quark fields $q$ and those without tilde to anti-quark fields $\bar{q}$. Quark contractions for the operator product $\mathcal{O}(x, y)\mathcal{O}^\dagger(x', y')$ may be specified by permutations from the set $\{\tilde{1}, \tilde{2}, \tilde{3}, \tilde{4}, \tilde{5}\}$ for quark fields to the set $\{1, 2, 3, 4, 5\}$ for anti-quark fields. For example, the contraction with crossed rectangular topology shown in Fig. 24 corresponds to

$$\begin{array}{ccccc} \tilde{1} & \tilde{2} & \tilde{3} & \tilde{4} & \tilde{5} \\ 4 & 5 & 2 & 3 & 1. \end{array} \tag{66}$$

There are 6, 18, 18 and 36 inequivalent contractions for the direct, crossed, rectangular and crossed rectangular diagrams. In Table 10 – 13 we summarize the weights of the corresponding amplitudes for $\pi$-$N$, $K$-$N$ and $\bar{K}$-$N$ four-point functions projected to isospin eigenchannels. Sign factors arising from Fermi statistics are not included in the weight. For definitions of meson and nucleon operators, see Sec. 2.2.

# Tables

Table 1: Experimental $s$-wave scattering lengths[1] and current algebra predictions in units of fm.

|         |           | experiment       | current algebra |
|---------|-----------|------------------|-----------------|
| $\pi$-$\pi$ | $I=0$     | $+0.37(7)$       | $+0.222$        |
|         | $I=2$     | $-0.040(17)$     | $-0.0635$       |
| $\pi$-$N$   | $I=1/2$   | $+0.245(4)$      | $+0.221$        |
|         | $I=3/2$   | $-0.143(6)$      | $-0.111$        |
| $N$-$N$     | $^3S_1$   | $-5.432(5)$      |                 |
|         | $^1S_0$   | $+20.1(4)$       |                 |
| $K$-$N$     | $I=0$     | $-0.0075$        | $0$             |
|         | $I=1$     | $-0.225$         | $-0.399$        |
| $\bar{K}$-$N$ | $I=0$   | $-1.16+i0.49$    | $+0.598$        |
|         | $I=1$     | $+0.17+i0.41$    | $+0.199$        |

Table 2: Parameters of simulation. All runs are made at $\beta = 5.7$ in quenched QCD.

|         | action | quark mass    | lattice size      | # conf. | gauge fixing     |
|---------|--------|---------------|-------------------|---------|------------------|
| $\pi$-$\pi$ | KS     | $m_q = 0.01$  | $8^3 \times 20$   | 400     | none             |
|         | KS     | $m_q = 0.01$  | $12^3 \times 20$  | 160     | none             |
|         | KS     | $m_q = 0.01$  | $12^3 \times 20$  | 60      | Coulomb          |
|         | Wilson | $K = 0.164$   | $12^3 \times 20$  | 70      | none             |
| $\pi$-$N$   | Wilson | $K = 0.164$   | $12^3 \times 20$  | 60      | $t=0$ Coulomb    |
|         | Wilson | $K = 0.1665$  | $12^3 \times 20$  | 30      | $t=0$ Coulomb    |
| $K$-$N$     | Wilson | $K = 0.164$   | $12^3 \times 20$  | 60      | $t=0$ Coulomb    |
| $\bar{K}$-$N$ | Wilson | $K = 0.164$ | $12^3 \times 20$  | 60      | $t=0$ Coulomb    |
| $N$-$N$     | Wilson | $K = 0.150$   | $20^3 \times 20$  | 20      | Coulomb          |
|         | Wilson | $K = 0.160$   | $20^3 \times 20$  | 30      | Coulomb          |
|         | Wilson | $K = 0.164$   | $20^3 \times 20$  | 20      | Coulomb          |

Table 3: Hadron masses and pion decay constant at $\beta = 5.7$ in quenched QCD. Also listed for comparison are results on a larger lattice from Refs. [23, 24]. For the Wilson result of Ref. [24] a linear interpolation in $1/K$ is made when necessary (marked by $*$).

| action | quark mass | lattice size | $m_\pi$ | $m_\rho$ | $m_N$ | $f_\pi$ |
|---|---|---|---|---|---|---|
| KS | $m_q = 0.01$ | $12^3 \times 20$ | 0.290(3) | | | 0.132(3) |
| | | $24^3 \times 32$[23] | 0.2876(7) | 0.883(58) | 1.454(26) | |
| Wilson | $K = 0.150$ | $12^3 \times 20$ | 1.0758(51) | 1.1302(66) | | 0.1900(12) |
| | | $20^3 \times 20$ | | | 1.788(11) | |
| | $K = 0.160$ | $12^3 \times 20$ | 0.6876(31) | 0.8053(39) | 1.2957(79) | 0.1268(14) |
| | | $20^3 \times 20$ | | | 1.302(13) | |
| | | $24^3 \times 32$[24] | 0.6887(11) | 0.8062(18) | 1.3068(51) | 0.1249(10) |
| | $K = 0.164$ | $12^3 \times 20$ | 0.5080(37) | 0.6865(53) | 1.080(10) | 0.1009(7) |
| | | $20^3 \times 20$ | | | 1.093(20) | |
| | | $24^3 \times 32$[24] | 0.5027(13)* | 0.6886(28)* | 1.0869(64)* | 0.1000(10)* |
| | $K = 0.1665$ | $12^3 \times 20$ | 0.3663(44) | 0.6086(75) | 0.926(13) | 0.0832(14) |
| | | $24^3 \times 32$[24] | 0.3656(17)* | 0.6165(45)* | 0.927(12)* | 0.0848(16)* |

Table 4: $\pi$-$\pi$ scattering lengths in lattice units for the Kogut-Susskind quark action at $m_q = 0.01$. Numbers in square brackets are current algebra predictions evaluated with the measured values of $m_\pi$ and $f_\pi$ listed in Table 3.

| size | gauge fixing | $I = 0$ | | | $I = 2$ | | |
|---|---|---|---|---|---|---|---|
| | | $a_0$ | $\delta E_I$ | $Z_I$ | $a_0$ | $\delta E_I$ | $Z_I$ |
| $8^3 \times 20$ | none | 4.89(53) | $-0.69(21)$ | 0.160(40) | $-0.374(14)$ | 0.0367(16) | 0.897(12) |
| $12^3 \times 20$ | none | 1.57(25) | $-0.0291(37)$ | 0.807(15) | $-0.301(28)$ | 0.00813(82) | 0.955(5) |
| | Coulomb | 1.73(27) | $-0.0316(40)$ | 0.800(17) | $-0.326(35)$ | 0.0089(10) | 0.948(7) |
| | | [1.16(5)] | | | [$-0.331(15)$] | | |

Table 5: $\pi$-$\pi$ scattering lengths in lattice units for the Wilson quark action at the hopping parameter $K = 0.164$ on a $12^3 \times 20$ lattice. Numbers in square brackets are current algebra predictions evaluated with the measured values of $m_\pi$ and $f_\pi$ listed in Table 3.

| | $I = 0$ | | | $I = 2$ | | |
|---|---|---|---|---|---|---|
| | $a_0$ | $\delta E_I$ | $Z_I$ | $a_0$ | $\delta E_I$ | $Z_I$ |
| | 3.02(17) | $-0.0297(19)$ | 0.903(7) | $-0.924(40)$ | 0.0166(9) | 1.027(5) |
| | [3.47(5)] | | | [$-0.993(16)$] | | |

Table 6: $\pi$-$N$ scattering lengths in lattice units for the Wilson quark action on a $12^3 \times 20$ lattice. Numbers in square brackets are current algebra predictions evaluated with the measured values of $m_\pi$, $m_N$ and $f_\pi$ listed in Table 3.

| | | $I = 1/2$ | | | $I = 3/2$ | | |
|---|---|---|---|---|---|---|---|
| $K$ | $a_0$ | $\delta E_I$ | $Z_I$ | $a_0$ | $\delta E_I$ | $Z_I$ |
| 0.164 | 3.04(66) | $-0.0219(54)$ | 0.916(22) | $-1.10(20)$ | 0.0151(35) | 0.975(18) |
| | [2.701(41)] | | | [$-1.350(20)$] | | |
| 0.1665 | $-0.70(58)$ | 0.011(11) | 1.021(52) | $-1.31(22)$ | 0.0243(55) | 0.964(31) |
| | [3.02(11)] | | | [$-1.509(53)$] | | |

Table 7: $K$-$N$ and $\bar{K}$-$N$ scattering lengths in lattice units for the Wilson quark action at the hopping parameter $K = 0.164$ on a $12^3 \times 20$ lattice. Numbers in square brackets are current algebra predictions evaluated with the measured values of $m_\pi$, $m_N$ and $f_\pi$ listed in Table 3.

| | | $I = 0$ | | | $I = 1$ | | |
|---|---|---|---|---|---|---|---|
| | $a_0$ | $\delta E_I$ | $Z_I$ | $a_0$ | $\delta E_I$ | $Z_I$ |
| $K - N$ | 0.55(47) | $-0.0051(38)$ | 0.987(19) | $-1.56(13)$ | 0.0240(27) | 1.012(16) |
| | [0] | | | [$-2.701(41)$] | | |
| $\bar{K} - N$ | 4.64(37) | $-0.0415(64)$ | 0.881(23) | 2.63(64) | $-0.0188(45)$ | 0.941(20) |
| | [4.051(61)] | | | [1.350(20)] | | |

Table 8: $N$-$N$ scattering lengths in lattice units for the Wilson quark action on a $20^4$ lattice.

| | | $^3S_1$ | | | $^1S_0$ | |
|---|---|---|---|---|---|---|
| $K$ | $a_0$ | $\delta E$ | $Z$ | $a_0$ | $\delta E$ | $Z$ |
| 0.150 | 10.8(1.2) | $-0.0126(39)$ | 1.012(19) | 9.2(1.3) | $-0.0085(27)$ | 0.987(13) |
| 0.160 | 9.0(1.6) | $-0.0109(45)$ | 1.048(25) | 7.3(1.9) | $-0.0072(33)$ | 1.007(18) |
| 0.164 | 10.8(9) | $-0.0207(48)$ | 1.021(12) | 8.0(1.1) | $-0.0102(30)$ | 0.997(7) |

Table 9: Exponent $\alpha$ governing the growth of error of the ratio $R^X(t)$ for a daigram of type $X$.

| $X$ | $D$ | $C$ | $R$ | $V$ | $CR$ |
|---|---|---|---|---|---|
| $\pi$-$\pi$ | 0 | 0 | $m_\pi$ | $2m_\pi$ | |
| $\pi$-$N$ | $m_N - \frac{3}{2}m_\pi$ | $m_N - \frac{3}{2}m_\pi$ | $m_N - \frac{1}{2}m_\pi$ | | $m_N - \frac{1}{2}m_\pi$ |
| $N$-$N$ | $2m_N - 3m_\pi$ | $2m_N - 3m_\pi$ | | | |

Table 10: Weights of direct(D) diagrams for $\pi$-$N$, $K$-$N$ and $\bar{K}$-$N$ four-point functions projected to isospin eigenchannels.

| | | $\pi$-$N$ | | $K$-$N$ | | $\bar{K}$-$N$ | |
|---|---|---|---|---|---|---|---|
| diagram | contraction | $I=1/2$ | $I=3/2$ | $I=0$ | $I=1$ | $I=0$ | $I=1$ |
| D1 | 12345 | 2 | 2 | 2 | 2 | 2 | 2 |
| D2 | 13245 | 1 | 1 | 1 | 1 | 1 | 1 |
| D3 | 21345 | $-2$ | $-2$ | $-2$ | $-2$ | $-2$ | $-2$ |
| D4 | 31245 | $-1$ | $-1$ | $-1$ | $-1$ | $-1$ | $-1$ |
| D5 | 32145 | 1 | 1 | 1 | 1 | 1 | 1 |
| D6 | 23145 | $-1$ | $-1$ | $-1$ | $-1$ | $-1$ | $-1$ |

Table 11: Weights of crossed(C) diagrams for $\pi$-$N$, $K$-$N$ and $\bar{K}$-$N$ four-point functions projected to isospin eigenchannels.

| diagram | contraction | $\pi$-$N$ $I = 1/2$ | $I = 3/2$ | $K$-$N$ $I = 0$ | $I = 1$ | $\bar{K}$-$N$ $I = 0$ | $I = 1$ |
|---|---|---|---|---|---|---|---|
| C1 | 42315 | 1 | 1 | 1 | 1 | 0 | 0 |
| C2 | 43215 | 3/2 | 0 | 2 | 0 | 0 | 0 |
| C3 | 24315 | −1 | −1 | −1 | −1 | 0 | 0 |
| C4 | 34215 | −3/2 | 0 | −2 | 0 | 0 | 0 |
| C5 | 23415 | 1/2 | −1 | 1 | −1 | 0 | 0 |
| C6 | 32415 | −1/2 | 1 | −1 | 1 | 0 | 0 |
| C7 | 41325 | −1 | −1 | −1 | −1 | 0 | 0 |
| C8 | 43125 | −3/2 | 0 | −2 | 0 | 0 | 0 |
| C9 | 14325 | 1 | 1 | 1 | 1 | 0 | 0 |
| C10 | 34125 | 3/2 | 0 | 2 | 0 | 0 | 0 |
| C11 | 13425 | −1/2 | 1 | −1 | 1 | 0 | 0 |
| C12 | 31425 | 1/2 | −1 | 1 | −1 | 0 | 0 |
| C13 | 41235 | 1/2 | −1 | 1 | −1 | 0 | 0 |
| C14 | 42135 | −1/2 | 1 | −1 | 1 | 0 | 0 |
| C15 | 14235 | −1/2 | 1 | −1 | 1 | 0 | 0 |
| C16 | 24135 | 1/2 | −1 | 1 | −1 | 0 | 0 |
| C17 | 12435 | −1 | 2 | −2 | 2 | 0 | 0 |
| C18 | 21435 | 1 | −2 | 2 | −2 | 0 | 0 |

Table 12: Weights of rectangular(R) diagrams for $\pi$-$N$, $K$-$N$ and $\bar{K}$-$N$ four-point functions projected to isospin eigenchannels.

| diagram | contraction | $\pi$-$N$ | | $K$-$N$ | | $\bar{K}$-$N$ | |
| | | $I = 1/2$ | $I = 3/2$ | $I = 0$ | $I = 1$ | $I = 0$ | $I = 1$ |
|---|---|---|---|---|---|---|---|
| R1 | 52341 | 1 | 1 | 0 | 0 | 1 | 1 |
| R2 | 53241 | $-1/2$ | 1 | 0 | 0 | $-1$ | 1 |
| R3 | 25341 | $-1$ | $-1$ | 0 | 0 | $-1$ | $-1$ |
| R4 | 35241 | $1/2$ | $-1$ | 0 | 0 | 1 | $-1$ |
| R5 | 23541 | $-3/2$ | 0 | 0 | 0 | $-2$ | 0 |
| R6 | 32541 | $3/2$ | 0 | 0 | 0 | 2 | 0 |
| R7 | 51342 | $-1$ | $-1$ | 0 | 0 | $-1$ | $-1$ |
| R8 | 53142 | $1/2$ | $-1$ | 0 | 0 | 1 | $-1$ |
| R9 | 15342 | 1 | 1 | 0 | 0 | 1 | 1 |
| R10 | 35142 | $-1/2$ | 1 | 0 | 0 | $-1$ | 1 |
| R11 | 13542 | $3/2$ | 0 | 0 | 0 | 2 | 0 |
| R12 | 31542 | $-3/2$ | 0 | 0 | 0 | $-2$ | 0 |
| R13 | 51243 | $-3/2$ | 0 | 0 | 0 | $-2$ | 0 |
| R14 | 52143 | $3/2$ | 0 | 0 | 0 | 2 | 0 |
| R15 | 15243 | $3/2$ | 0 | 0 | 0 | 2 | 0 |
| R16 | 25143 | $-3/2$ | 0 | 0 | 0 | $-2$ | 0 |
| R17 | 12543 | 3 | 0 | 0 | 0 | 4 | 0 |
| R18 | 21543 | $-3$ | 0 | 0 | 0 | $-4$ | 0 |

Table 13: Weights of crossed rectangular(CR) diagrams for $\pi$-$N$, $K$-$N$ and $\bar{K}$-$N$ four-point functions projected to isospin eigenchannels.

| | | $\pi$-$N$ | | $K$-$N$ | | $\bar{K}$-$N$ | |
|---|---|---|---|---|---|---|---|
| diagram | contraction | $I = 1/2$ | $I = 3/2$ | $I = 0$ | $I = 1$ | $I = 0$ | $I = 1$ |
| CR1 | 52431 | $-1/2$ | 1 | 0 | 0 | 0 | 0 |
| CR2 | 54231 | $-1/2$ | 1 | 0 | 0 | 0 | 0 |
| CR3 | 53421 | $-1/2$ | 1 | 0 | 0 | 0 | 0 |
| CR4 | 54321 | 1 | 1 | 0 | 0 | 0 | 0 |
| CR5 | 25431 | $1/2$ | $-1$ | 0 | 0 | 0 | 0 |
| CR6 | 45231 | $1/2$ | $-1$ | 0 | 0 | 0 | 0 |
| CR7 | 35421 | $1/2$ | $-1$ | 0 | 0 | 0 | 0 |
| CR8 | 45321 | $-1$ | $-1$ | 0 | 0 | 0 | 0 |
| CR9 | 24531 | 0 | 0 | 0 | 0 | 0 | 0 |
| CR10 | 42531 | 0 | 0 | 0 | 0 | 0 | 0 |
| CR11 | 34521 | $3/2$ | 0 | 0 | 0 | 0 | 0 |
| CR12 | 43521 | $-3/2$ | 0 | 0 | 0 | 0 | 0 |
| CR13 | 51432 | $1/2$ | $-1$ | 0 | 0 | 0 | 0 |
| CR14 | 54132 | $1/2$ | $-1$ | 0 | 0 | 0 | 0 |
| CR15 | 53412 | $1/2$ | $-1$ | 0 | 0 | 0 | 0 |
| CR16 | 54312 | $-1$ | $-1$ | 0 | 0 | 0 | 0 |
| CR17 | 15432 | $-1/2$ | 1 | 0 | 0 | 0 | 0 |
| CR18 | 45132 | $-1/2$ | 1 | 0 | 0 | 0 | 0 |
| CR19 | 35412 | $-1/2$ | 1 | 0 | 0 | 0 | 0 |
| CR20 | 45312 | 1 | 1 | 0 | 0 | 0 | 0 |
| CR21 | 14532 | 0 | 0 | 0 | 0 | 0 | 0 |
| CR22 | 41532 | 0 | 0 | 0 | 0 | 0 | 0 |
| CR23 | 34512 | $-3/2$ | 0 | 0 | 0 | 0 | 0 |
| CR24 | 43512 | $3/2$ | 0 | 0 | 0 | 0 | 0 |
| CR25 | 51423 | 0 | 0 | 0 | 0 | 0 | 0 |
| CR26 | 54123 | $3/2$ | 0 | 0 | 0 | 0 | 0 |
| CR27 | 52413 | 0 | 0 | 0 | 0 | 0 | 0 |
| CR28 | 54213 | $-3/2$ | 0 | 0 | 0 | 0 | 0 |
| CR29 | 15423 | 0 | 0 | 0 | 0 | 0 | 0 |
| CR30 | 45123 | $-3/2$ | 0 | 0 | 0 | 0 | 0 |
| CR31 | 25413 | 0 | 0 | 0 | 0 | 0 | 0 |
| CR32 | 45213 | $3/2$ | 0 | 0 | 0 | 0 | 0 |
| CR33 | 14523 | $3/2$ | 0 | 0 | 0 | 0 | 0 |
| CR34 | 41523 | $-3/2$ | 0 | 0 | 0 | 0 | 0 |
| CR35 | 24513 | $-3/2$ | 0 | 0 | 0 | 0 | 0 |
| CR36 | 42513 | $3/2$ | 0 | 0 | 0 | 0 | 0 |

# Figure Captions

Figure 1: Diagrams contributing to $\pi$-$\pi$ four-point functions. Short bars represent wall sources. Open circles are sinks for local pion operator.

Figure 2: Diagrams contributing to $\pi$-$N$ four-point functions. Short bars represent wall sources. Open circles are sinks for local pion or nucleon operator. Solid circle for the $CR$ diagram means use of source method for calculating the product of quark propagators connected by it.

Figure 3: Diagrams contributing to $N$-$N$ four-point functions. Short bars represent wall sources. Open circles are sinks for local nucleon operator.

Figure 4: Individual ratios $R^X(t)$ ($X = D, C, R$ and $V$) for $\pi$-$\pi$ four-point function calculated with the method of wall source without gauge fixing for Kogut-Susskind quark action on a $12^3 \times 20$ lattice at $\beta = 5.7$ and $m_q = 0.01$.

Figure 5: $R_I(t)$ ($I = 0$ and 2) for $\pi$-$\pi$ four-point function calculated without gauge fixing for the Kogut-Susskind quark action on a $12^3 \times 20$ lattice at $\beta = 5.7$ and $m_q = 0.01$. Solid lines are linear fits for $4 \leq t \leq 9$.

Figure 6: $R_I(t)$ ($I = 0$ and 2) for $\pi$-$\pi$ four-point function calculated without gauge fixing for the Kogut-Susskind quark action on an $8^3 \times 20$ lattice at $\beta = 5.7$ and $m_q = 0.01$. Solid lines are linear fits for $4 \leq t \leq 9$.

Figure 7: Comparison of $R_I(t)$ ($I = 0$ and 2) for $\pi$-$\pi$ four-point function calculated in Coulomb gauge (open symbols) and without gauge fixing (filled symbols) for the Kogut-Susskind quark action for the same set of 60 configurations on a $12^3 \times 20$ lattice at $\beta = 5.7$ and $m_q = 0.01$.

Figure 8: Comparison of single-elimination jackknife errors of $R_I(t)$ ($I = 0$ and 2) for $\pi$-$\pi$ four-point function calculated in Coulomb gauge (open symbols) and without gauge fixing (filled symbols) for the Kogut-Susskind quark action for the same set of 60 configurations on a $12^3 \times 20$ lattice at $\beta = 5.7$ and $m_q = 0.01$.

Figure 9: Individual ratios $R^X(t)$ ($X = D, C, R$ and $V$) for $\pi$-$\pi$ four-point function calclulated without gauge fixing for Wilson quark action on a $12^3 \times 20$ lattice at $\beta = 5.7$ and $K = 0.164$.

Figure 10: $R_I(t)$ ($I = 0$ and 2) for $\pi$-$\pi$ four-point function calculated without gauge fixing for the Wilson quark action on a $12^3 \times 20$ lattice at $\beta = 5.7$ and $K = 0.164$. Solid lines are linear fits for $4 \leq t \leq 9$.

Figure 11: Fractional change of the $I = 0$ $\pi$-$\pi$ scattering length due to infrared divergences in quenched QCD normalized by the current algebra value $a_0 = 7m_\pi/(32\pi f_\pi^2)$ at $\beta = 5.7$ plotted as a function of $m_\pi/m_\rho$. Solid line is for the Wilson action and dotted line for the Kogut-Susskind action. Arrows indicate the position where our calculation of $\pi$-$\pi$ four-point functions is made.

Figure 12: $I = 0$ and 2 $s$-wave $\pi$-$\pi$ scattering lengths $a_0^I$. Filled and open symbols denote Wilson and Kogut-Susskind results. Triangles are for Coulomb gauge results. Squares ($\beta = 5.7$) and diamonds ($\beta = 6.0$) for $I = 2$ are from Refs. [18, 19]. Dotted lines indicate predictions of current algebra.

Figure 13: $R_I(t)$ ($I = 1/2$ and 3/2) for the $\pi$-$N$ four-point function at (a) $K = 0.164$ and (b) 0.1665. Solid lines are linear fits for $4 \leq t \leq 9$.

Figure 14: Individual ratios $R_\pm^X(t)$ ($X = D, C, R$ and $CR$) for $\pi$-$N$ four-point function projected to crossing even (filled circles) and odd (open circles) channels.

Figure 15: Individual ratios $R_I^X(t)$ ($X = D$ and $C$) of the $K$-$N$ amplitude for isospin eigenchannels. Filled and open symbols denote $I = 0$ and 1 results.

Figure 16: Individual ratios $R_I^X(t)$ ($X = D$ and $R$) of the $\bar{K}$-$N$ amplitude for isospin eigenchannels. Filled and open symbols denote $I = 0$ and 1 results.

Figure 17: $R_I(t)$ ($I = 0$ and 1) for the $K$-$N$ four-point function on a $12^3 \times 20$ lattice at $\beta = 5.7$ and $K = 0.164$. Solid lines are linear fits for $4 \leq t \leq 9$.

Figure 18: $R_I(t)$ ($I = 0$ and 1) for the $\bar{K}$-$N$ four-point function on a $12^3 \times 20$ lattice at $\beta = 5.7$ and $K = 0.164$. Solid lines are linear fits for $4 \leq t \leq 9$.

Figure 19: Individual contributions of direct and crossed amplitudes to the ratio $R(t)$ for triplet $^3S_1$(filled symbols) and singlet $^1S_0$(open symbols) channels.

Figure 20: $R(t)$ for $N$-$N$ four point function on a $20^4$ lattice at $\beta = 5.7$ and $K = 0.160$. Solid lines are linear fits for $4 \leq t \leq 9$.

Figure 21: $N$-$N$ scattering lengths in units of fm as compared to $\pi$-$N$ and $\pi$-$\pi$ scattering lenghts in quenched QCD at $\beta = 5.7$ calculated with the Wilson quark action. Conversion to physical units is made with $a = 0.137(2)$fm determined from the $\rho$ meson mass.

Figure 22: Error $\delta R^X(t)$ of ratio $R^X(t)$ as a function of $t$ at $\beta = 5.7$ in qenched QCD. Use of Kogut-Susslind or Wilson quark action is indicated in parentheses. Solid lines indicate expected slope calculated with measured values of $m_N$ and $m_\pi$. (a) $\pi$-$\pi$ at $m_q = 0.01$, (b) $\pi$-$\pi$ at $K = 0.164$, (c) $\pi$-$N$ at $K = 0.164$, (d) $N$-$N$ in the triplet $^3S_1$ channel at $K = 0.160$.

Figure 23: Largest exponent $\alpha$ governing the errors of ratio $\delta R(t) \propto e^{\alpha t}$ for $\pi$-$\pi$ (dotted line), $\pi$-$N$ (broken line) and $N$-$N$ (solid line) cases.

Figure 24: Example of quark contractions contributing to $\pi$-$N$ four-point function.

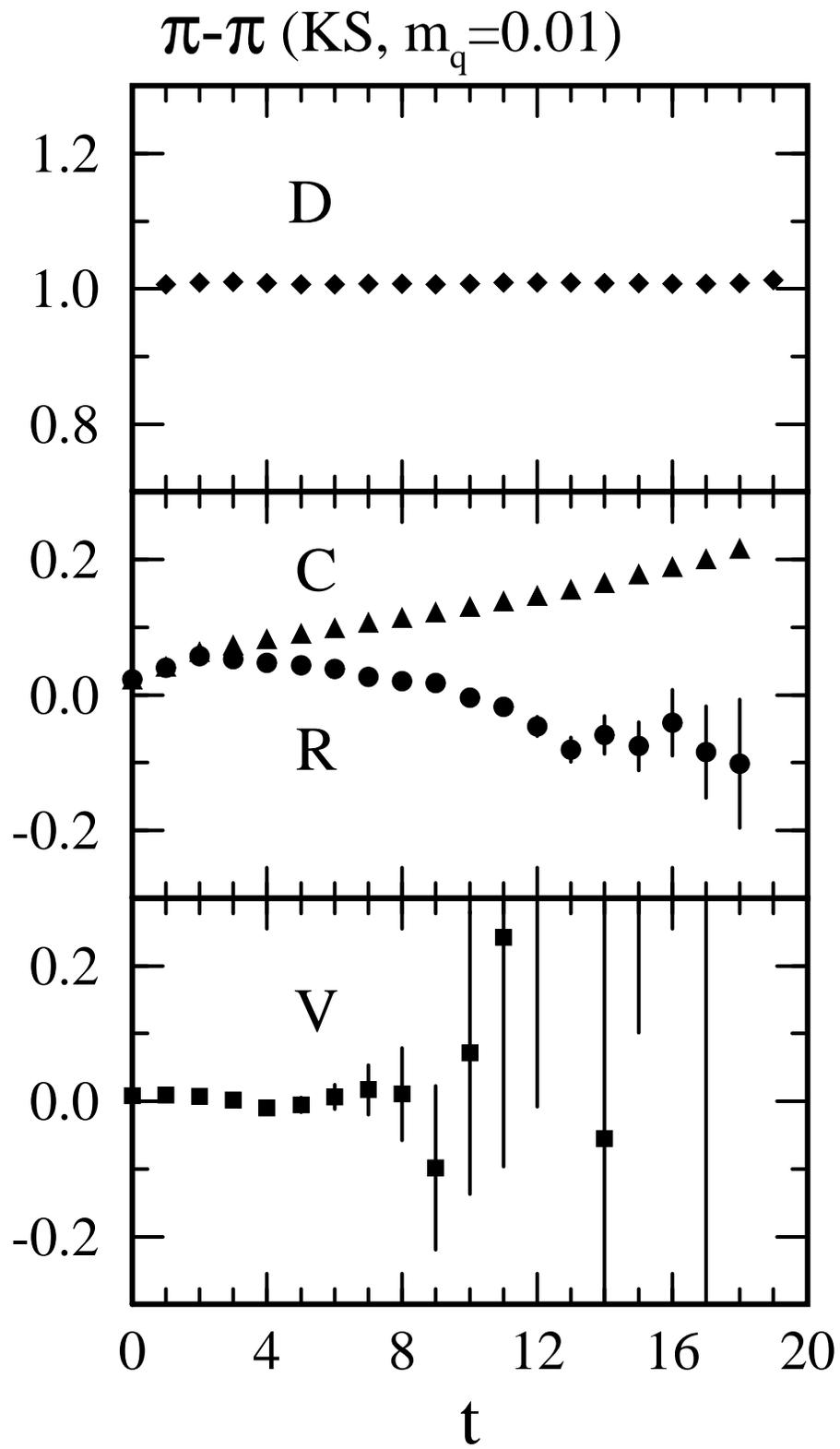

$\pi$-$\pi$ (KS, $m_q$=0.01)

Fig.4

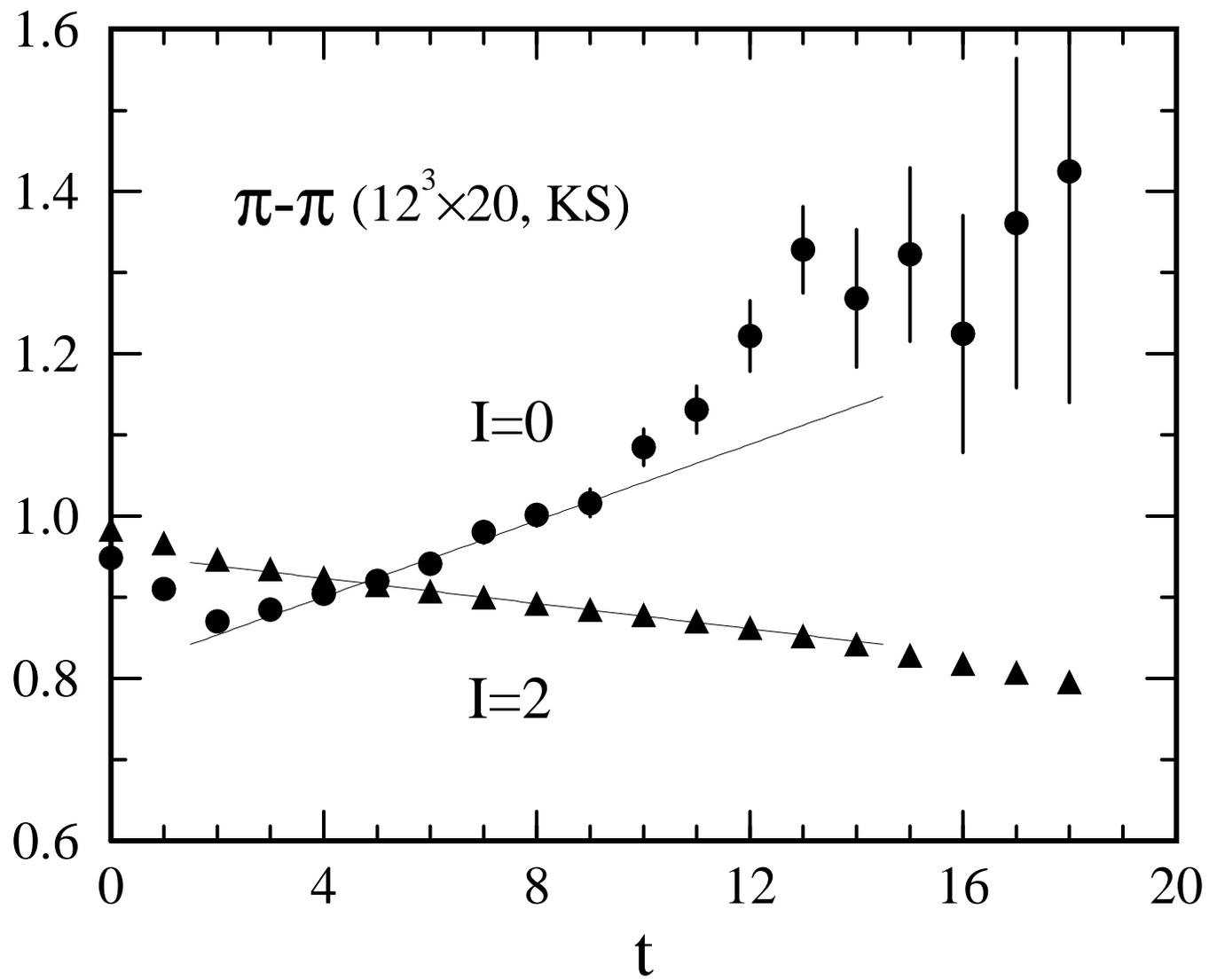

Fig.5

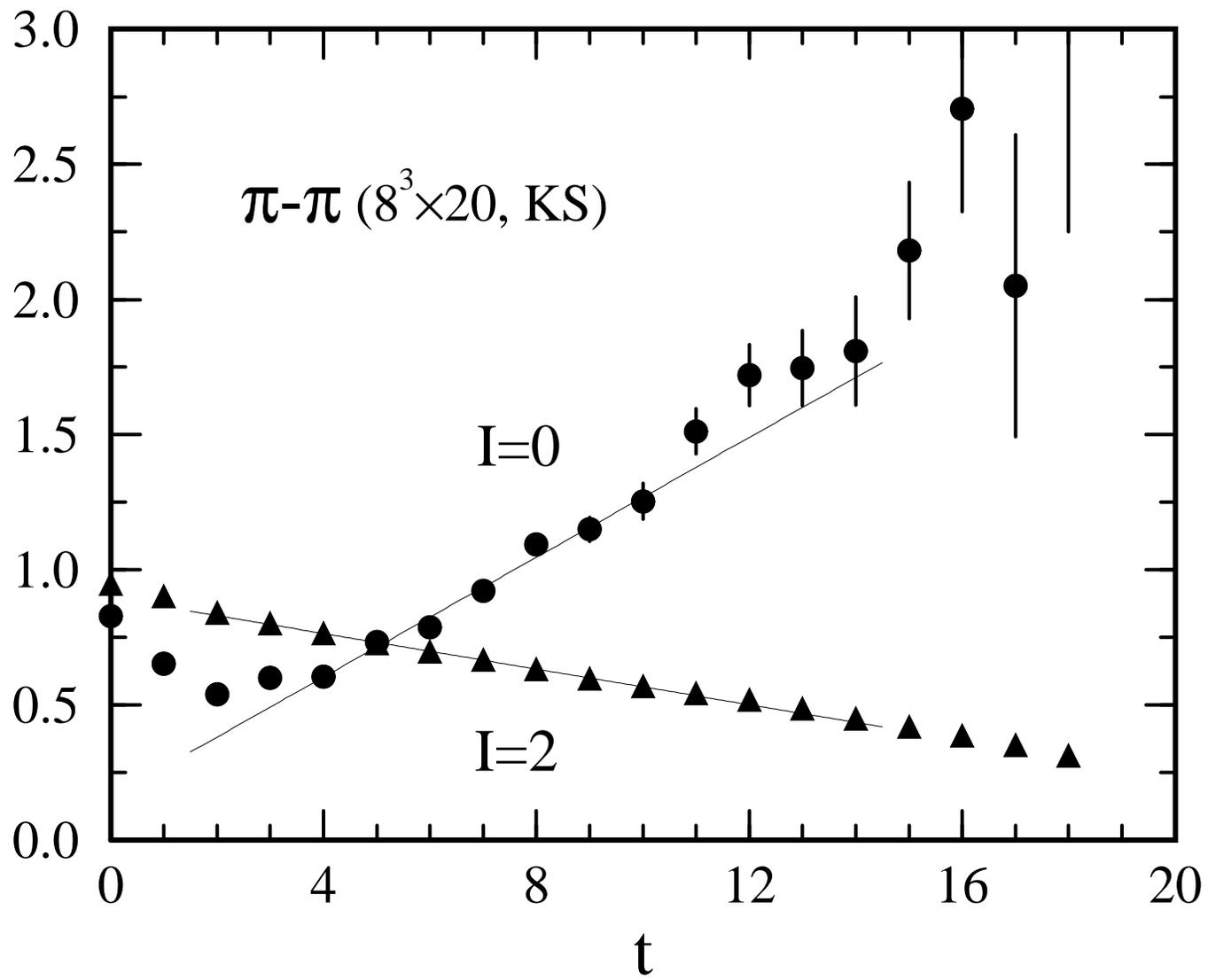

Fig.6

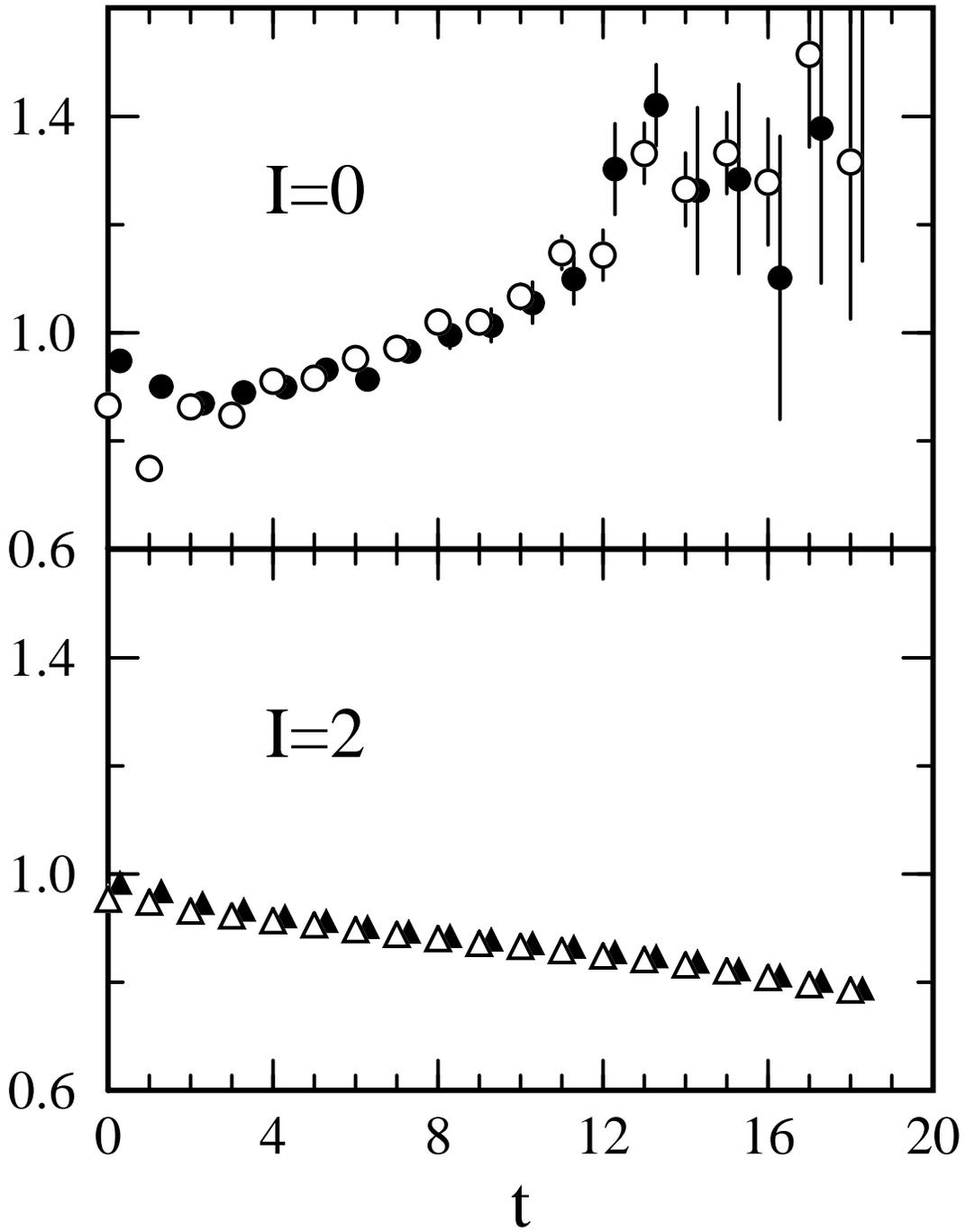

Fig.7

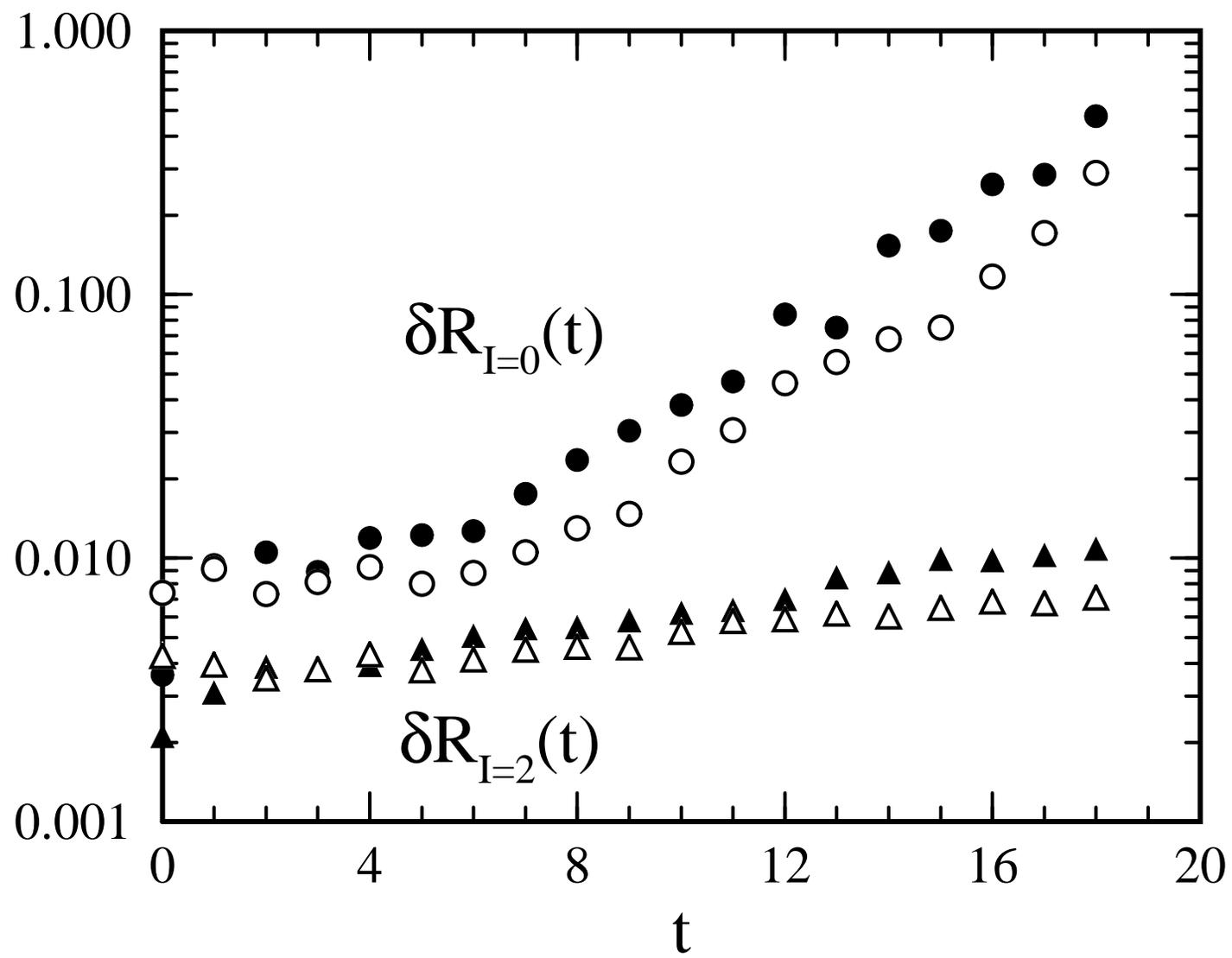

$\delta R_{I=0}(t)$

$\delta R_{I=2}(t)$

t

Fig.8

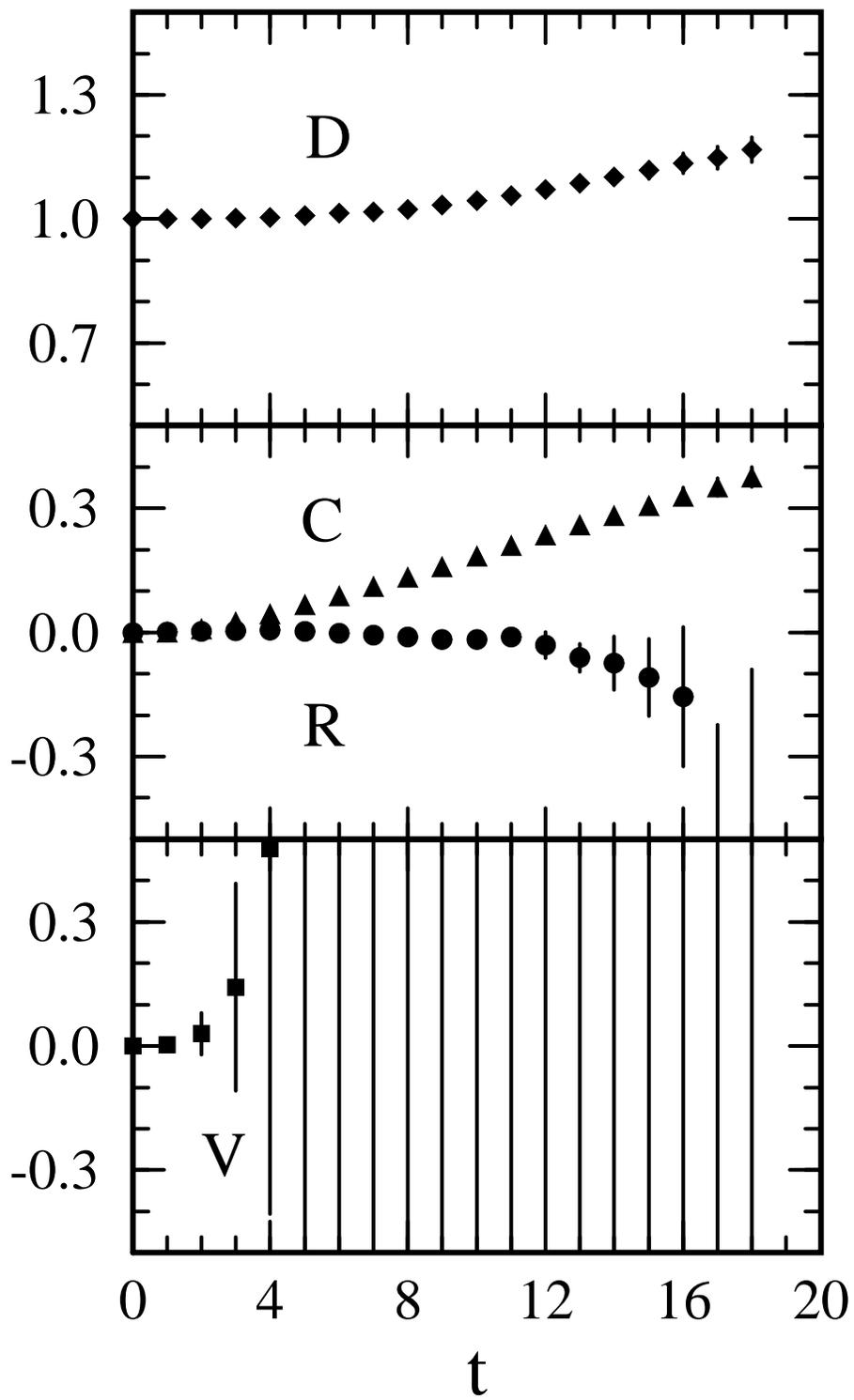

π-π (Wilson, K=0.164)

Fig.9

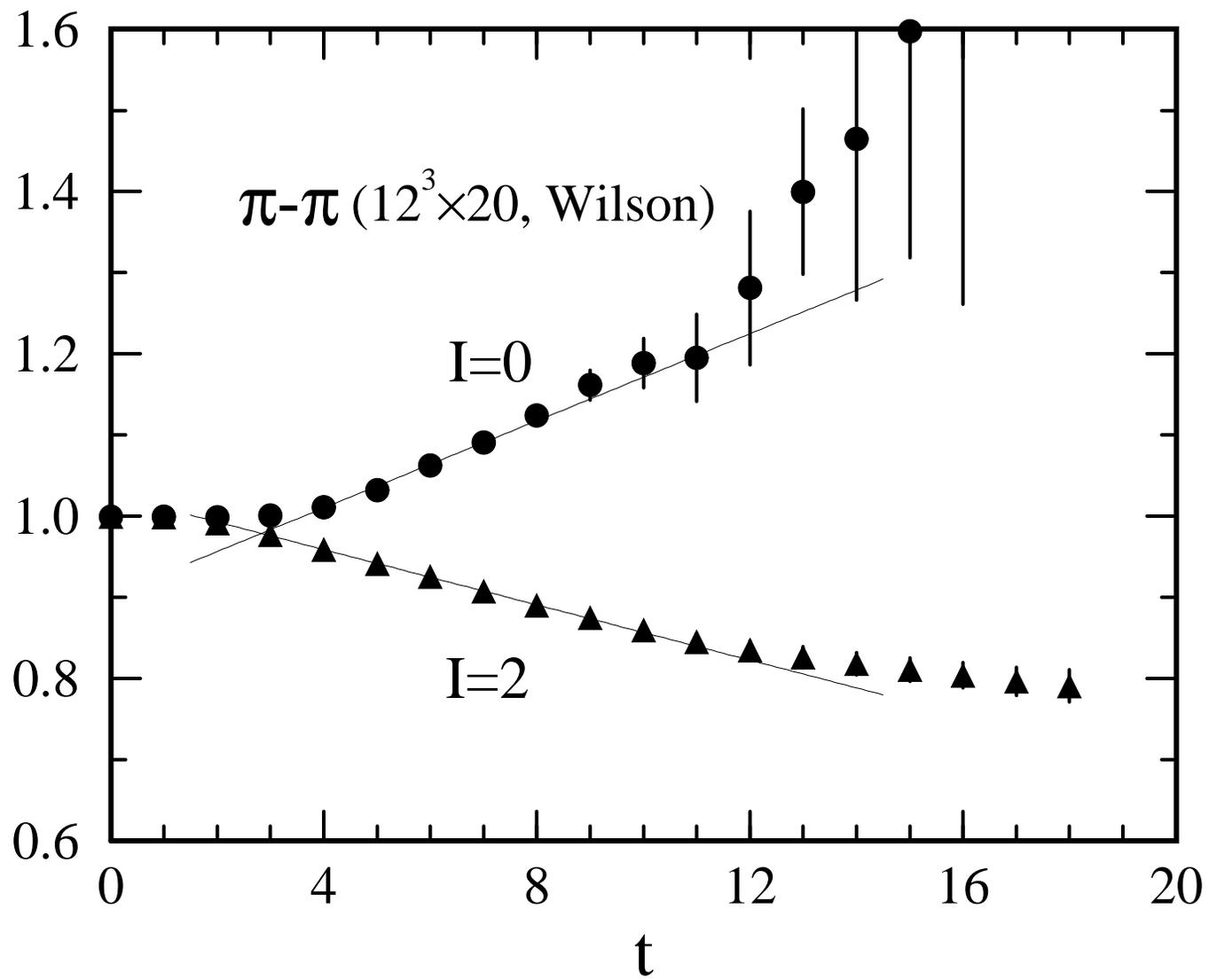

Fig.10

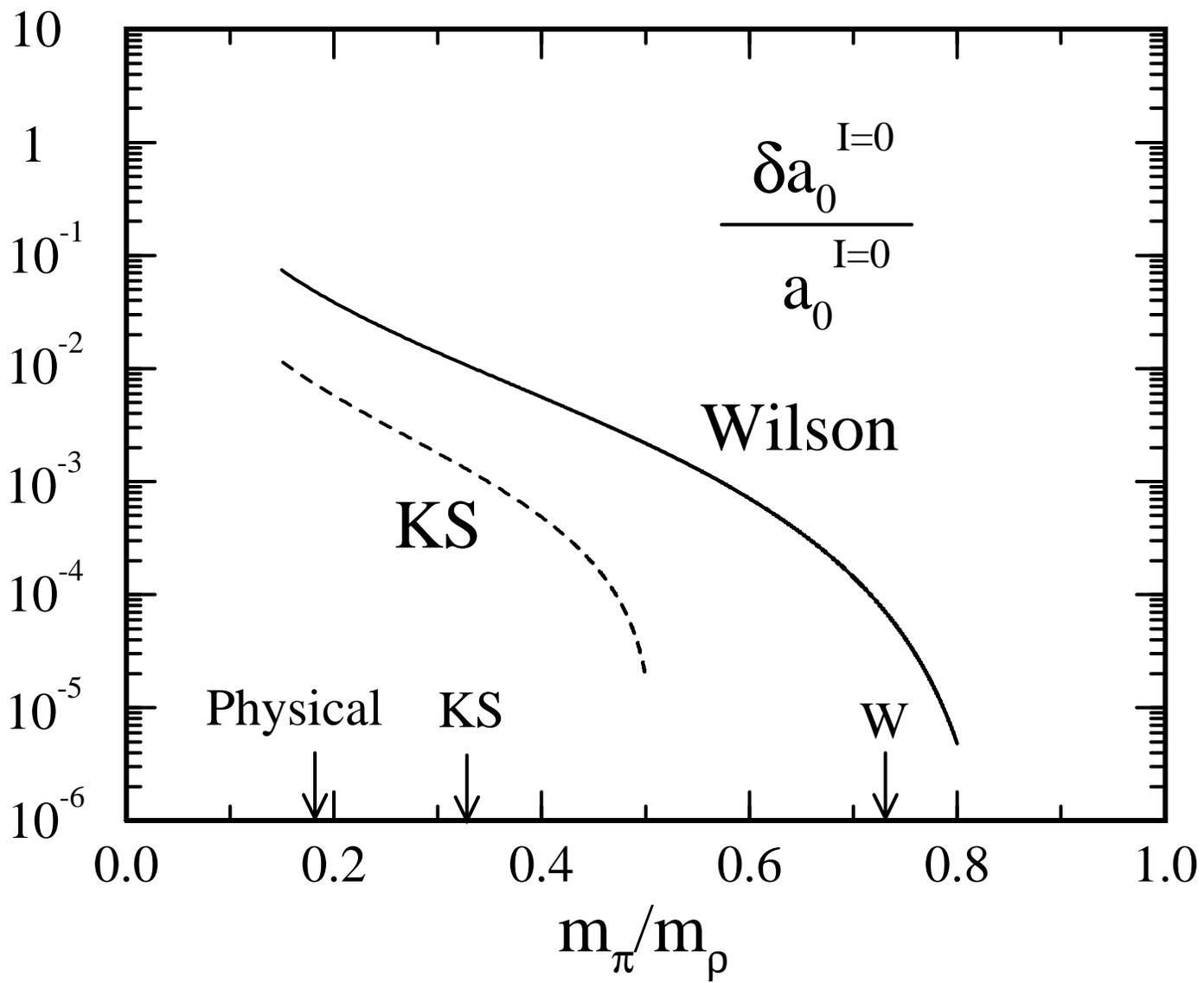

Fig.11

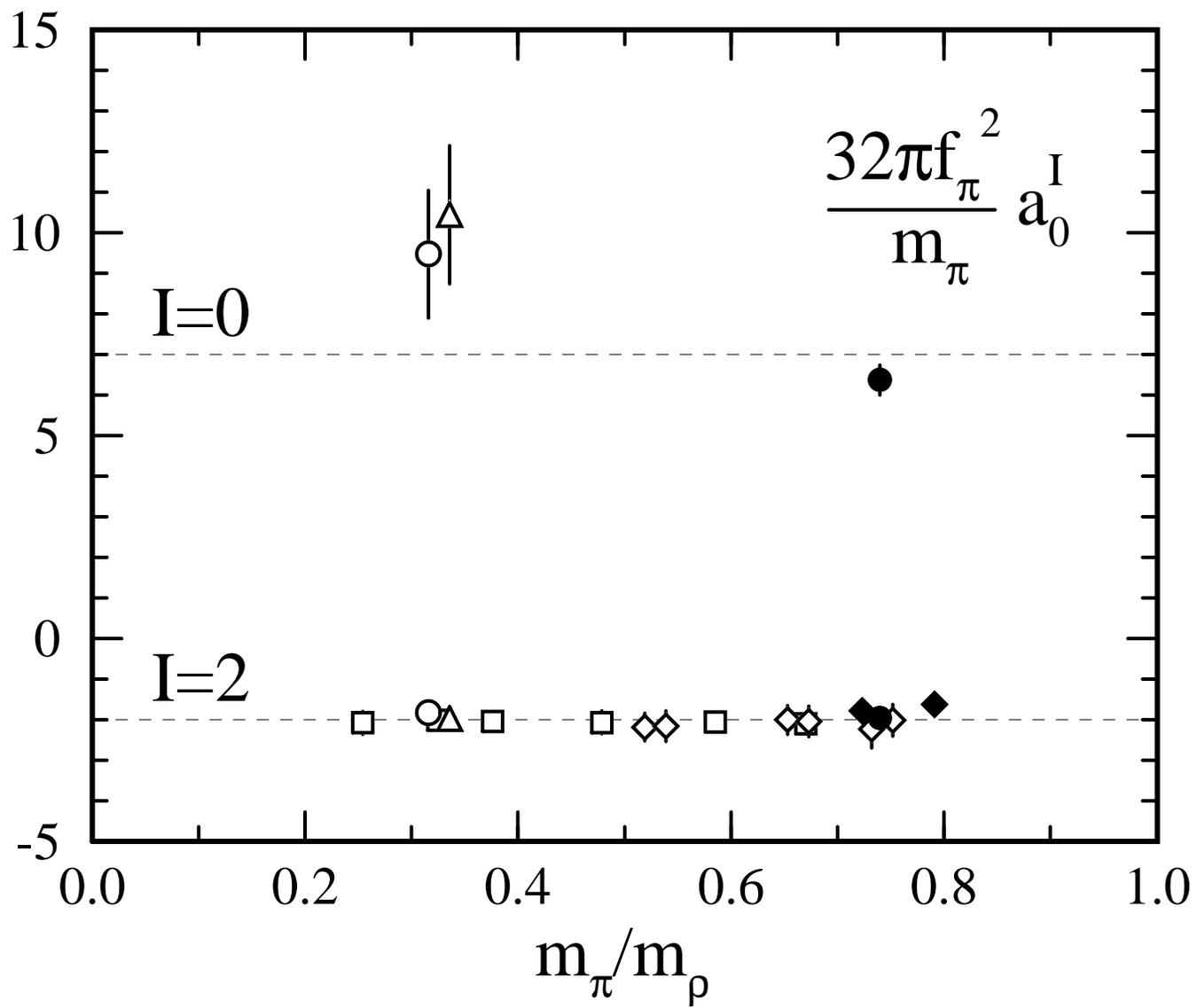

Fig.12

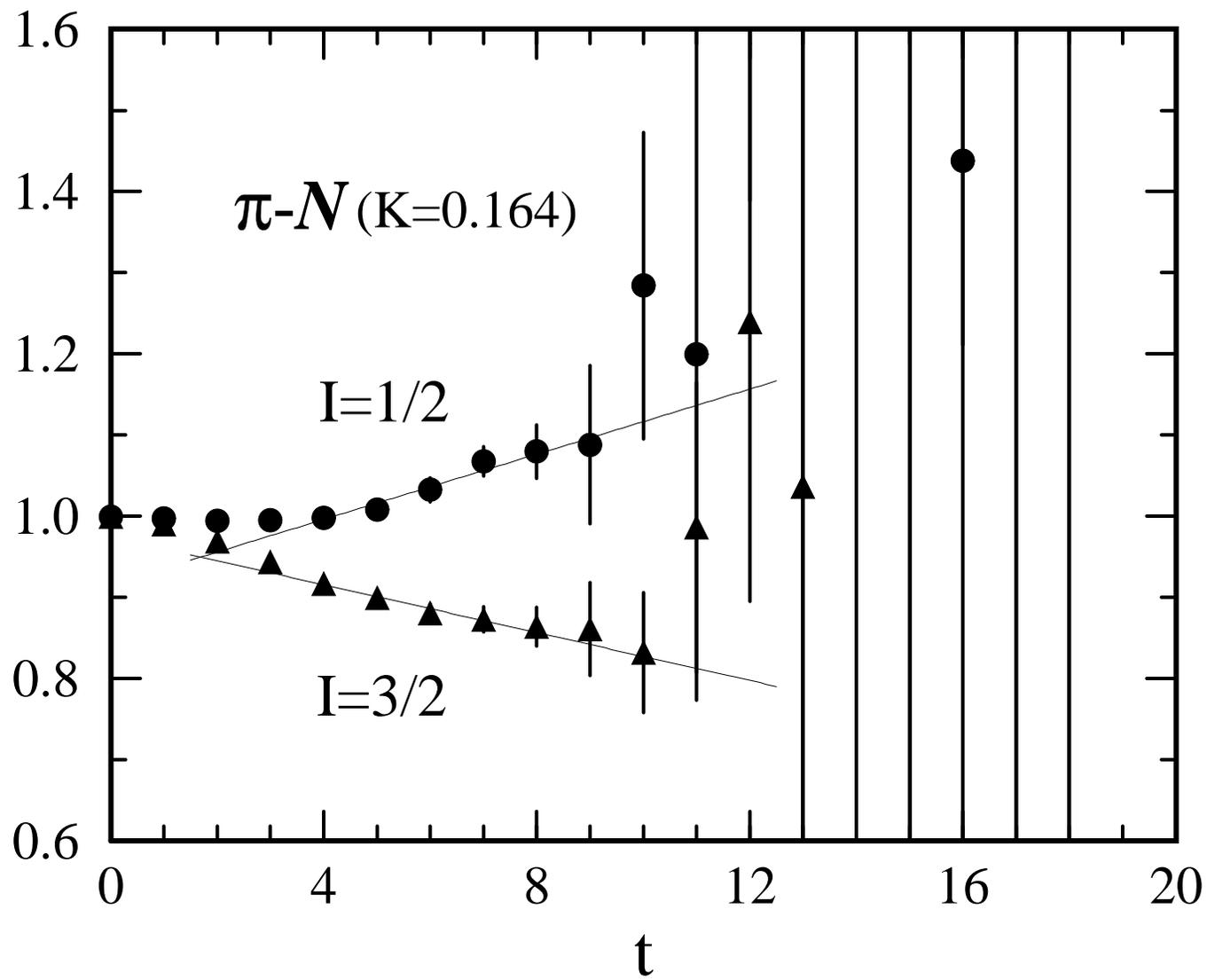

Fig.13(a)

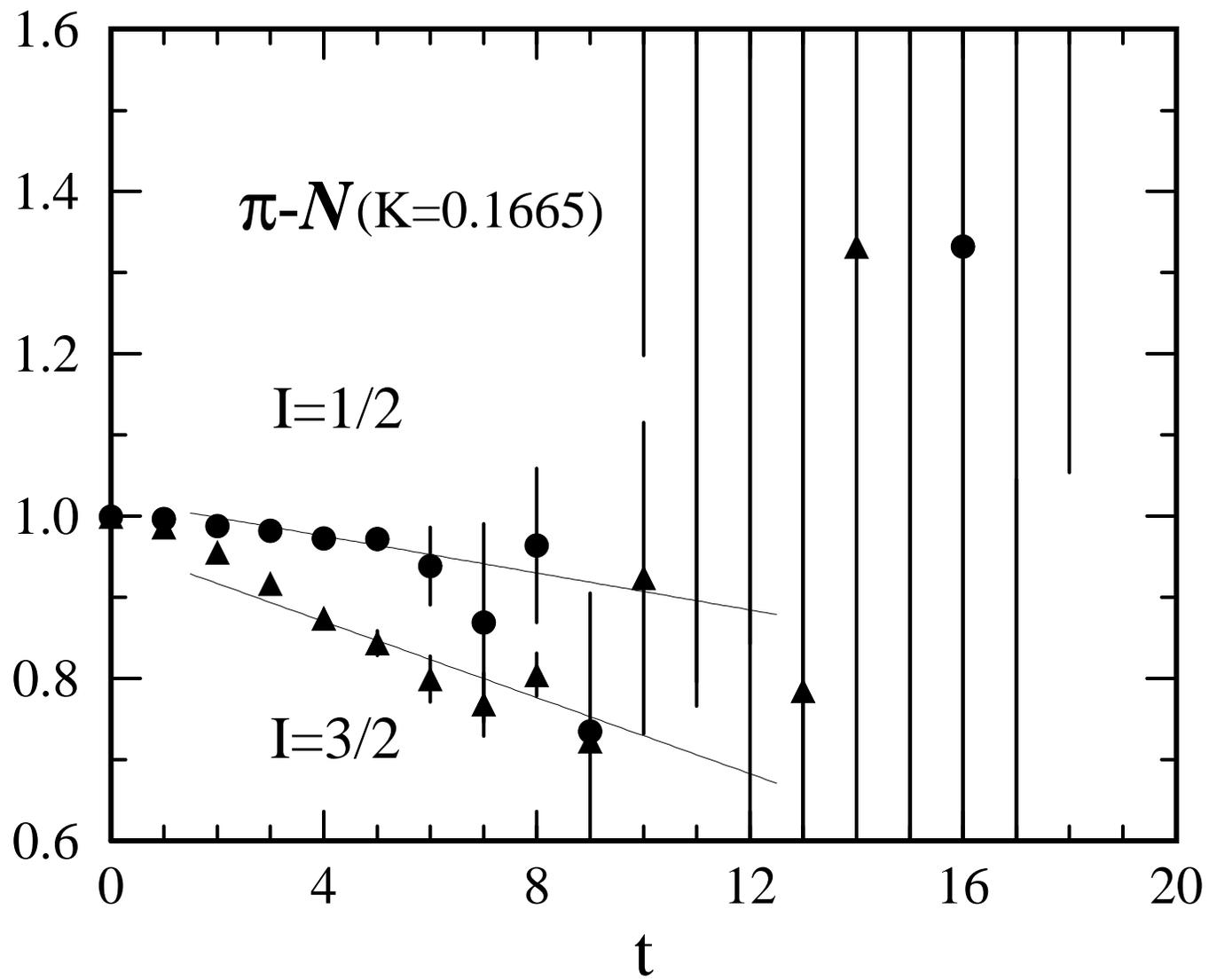

$\pi$-$N$ (K=0.1665)

I=1/2

I=3/2

t

Fig.13(b)

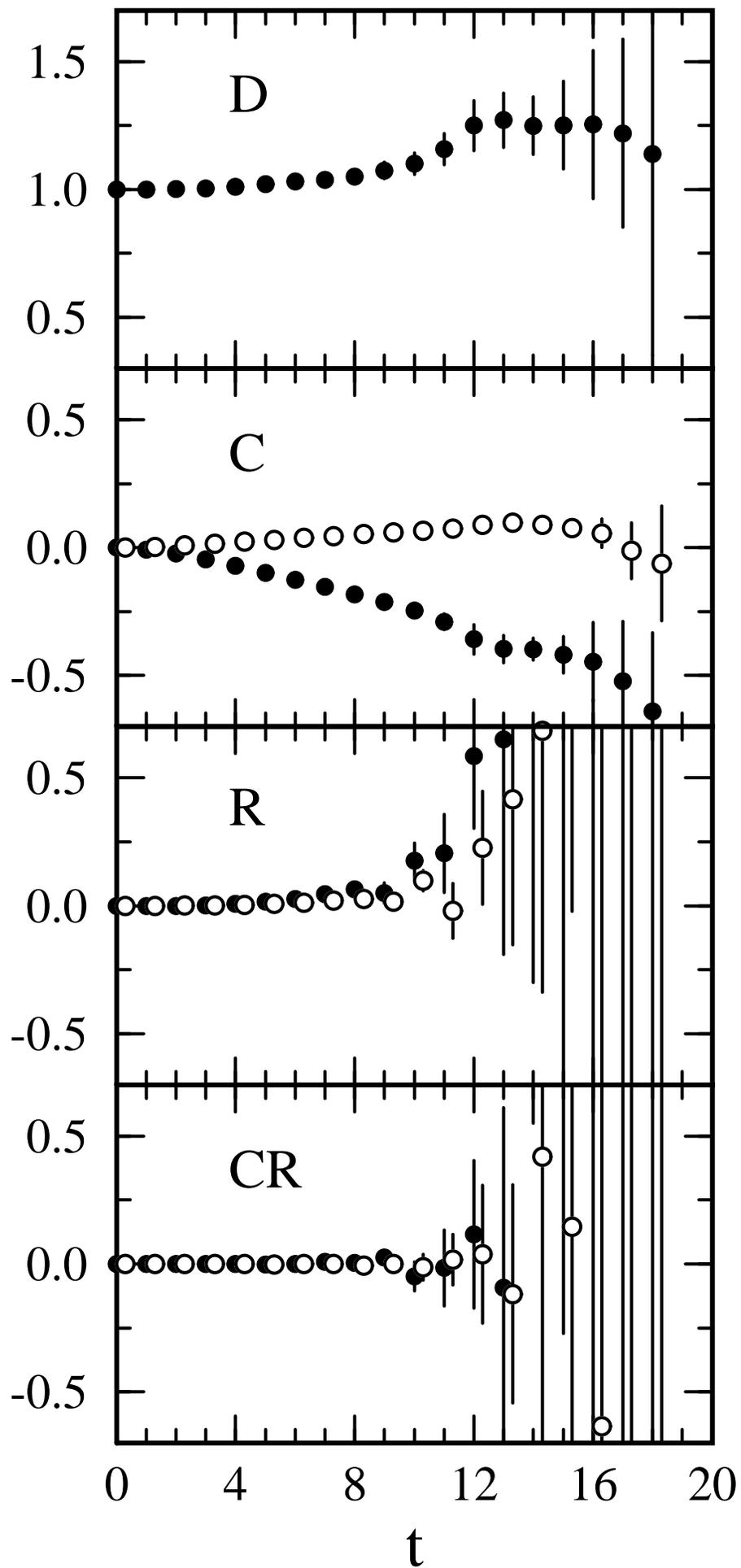

$\pi$-$N$ (Wilson, K=0.164)

Fig.14

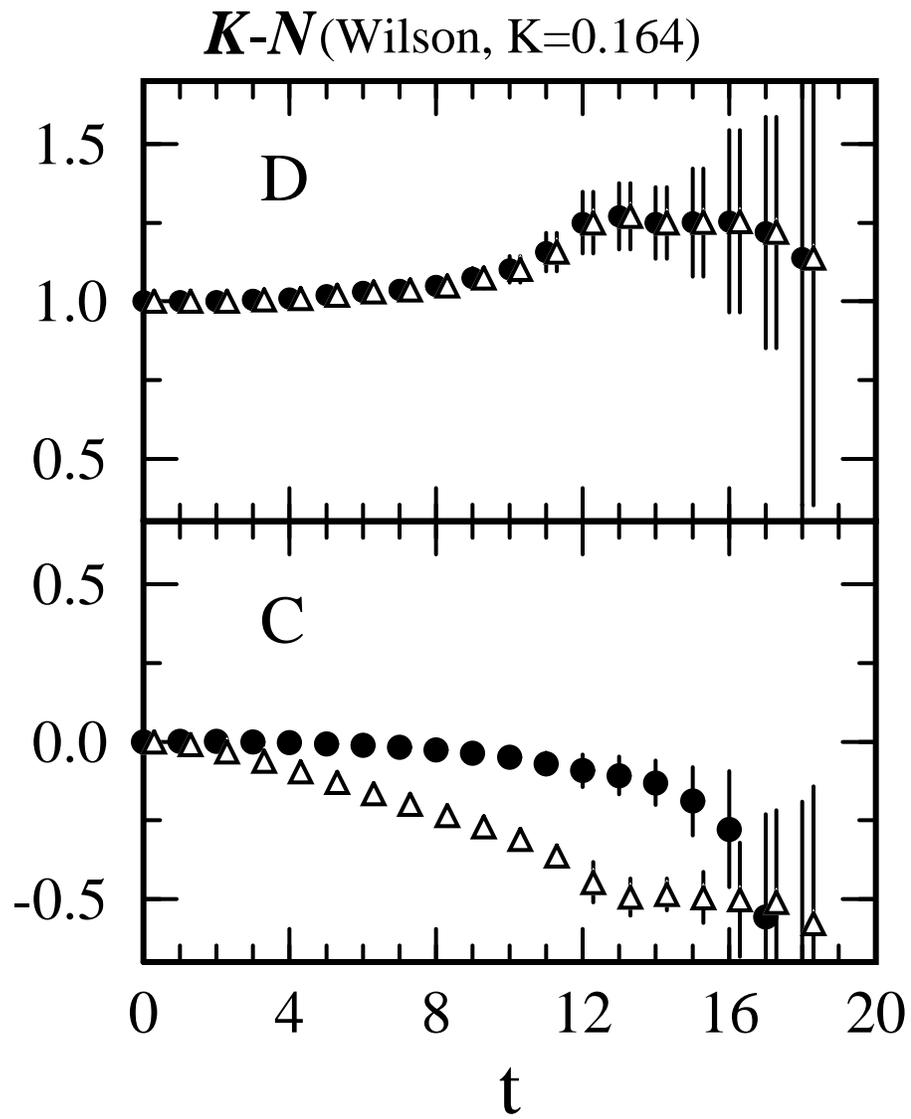

**K-N**(Wilson, K=0.164)

Fig.15

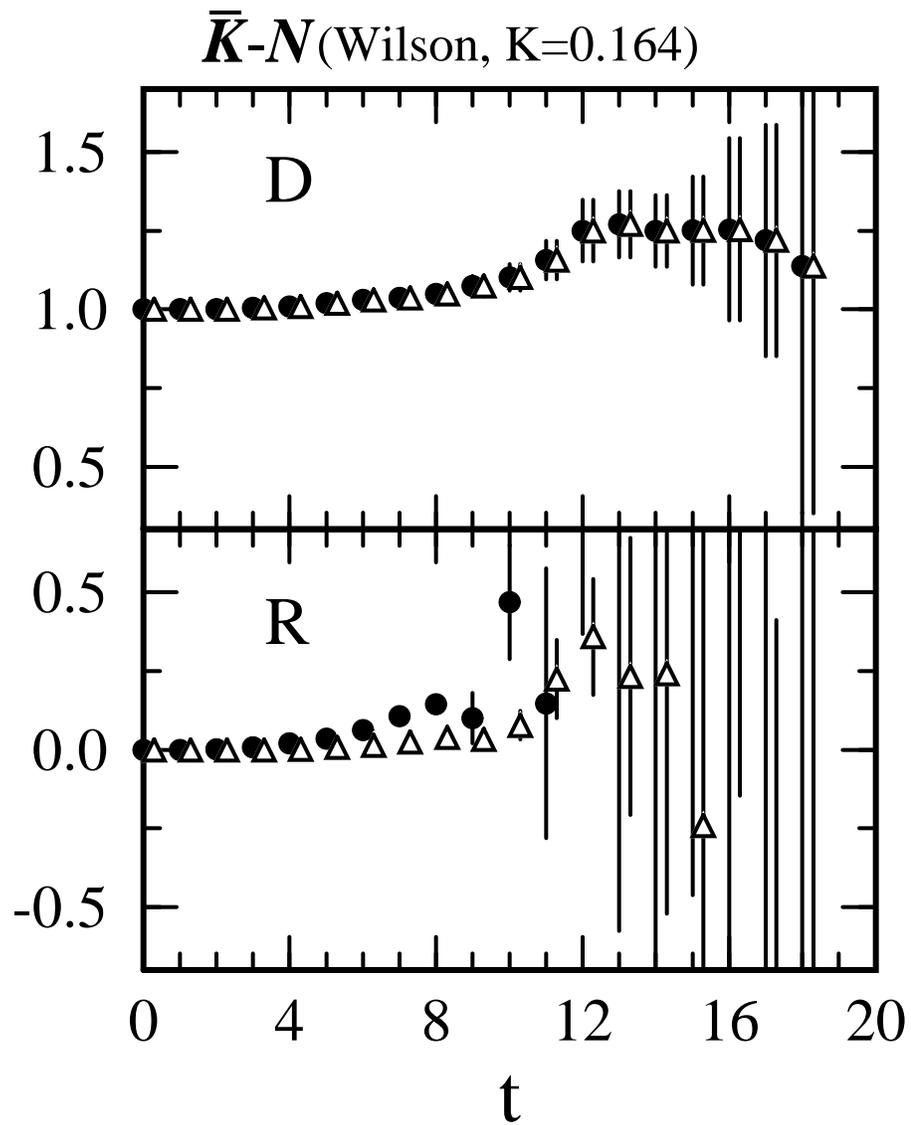

$\bar{K}$-$N$(Wilson, K=0.164)

Fig.16

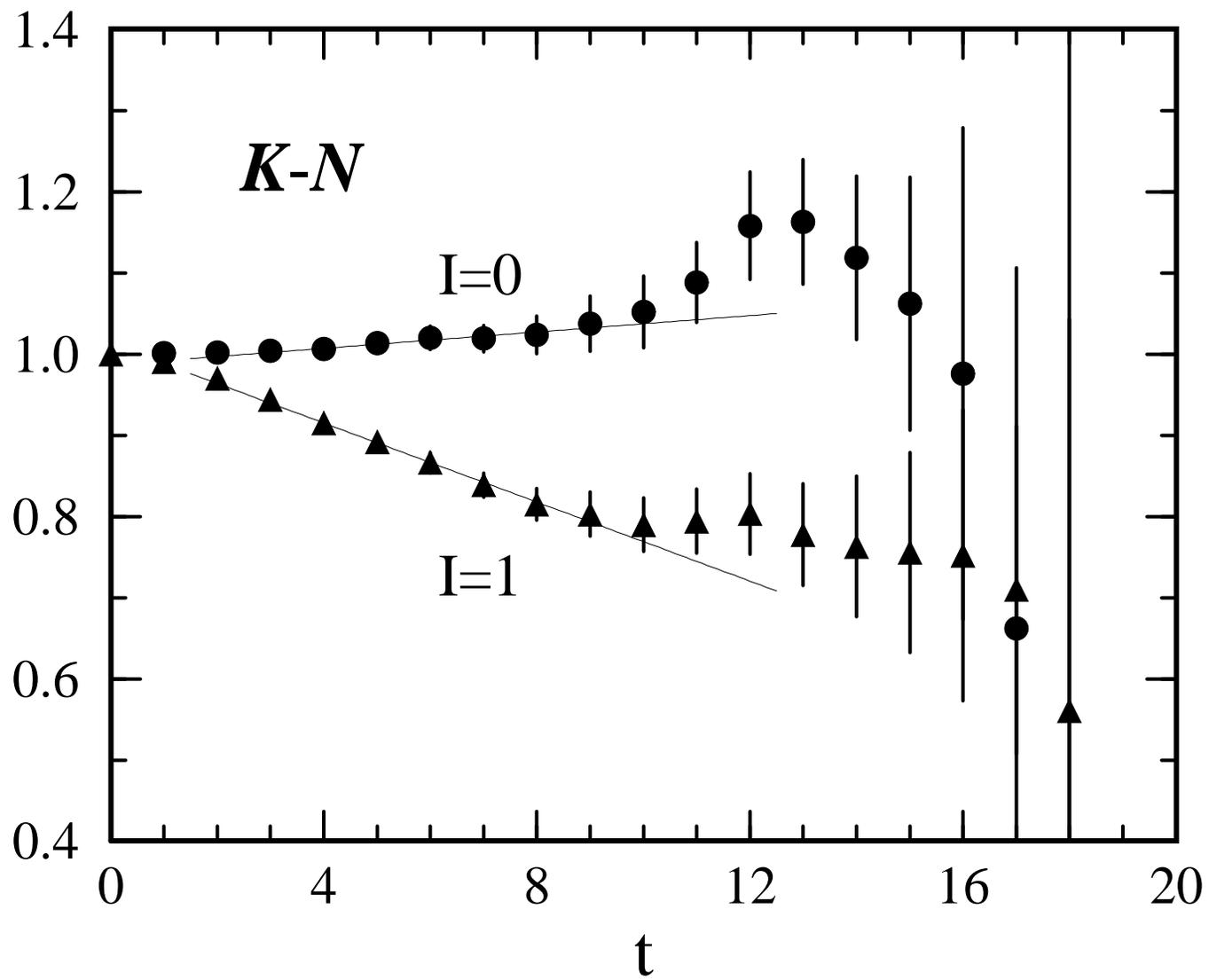

Fig.17

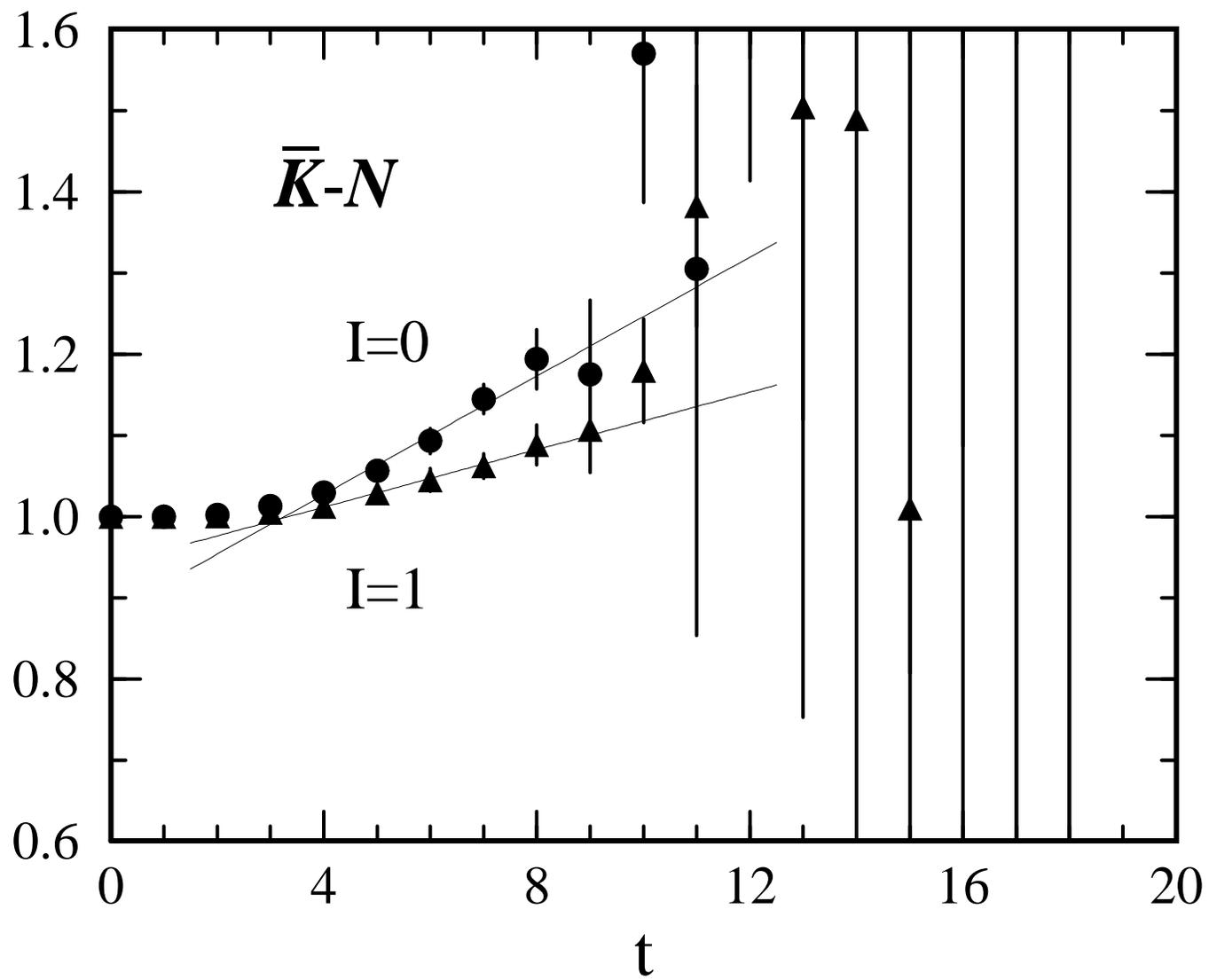

Fig.18

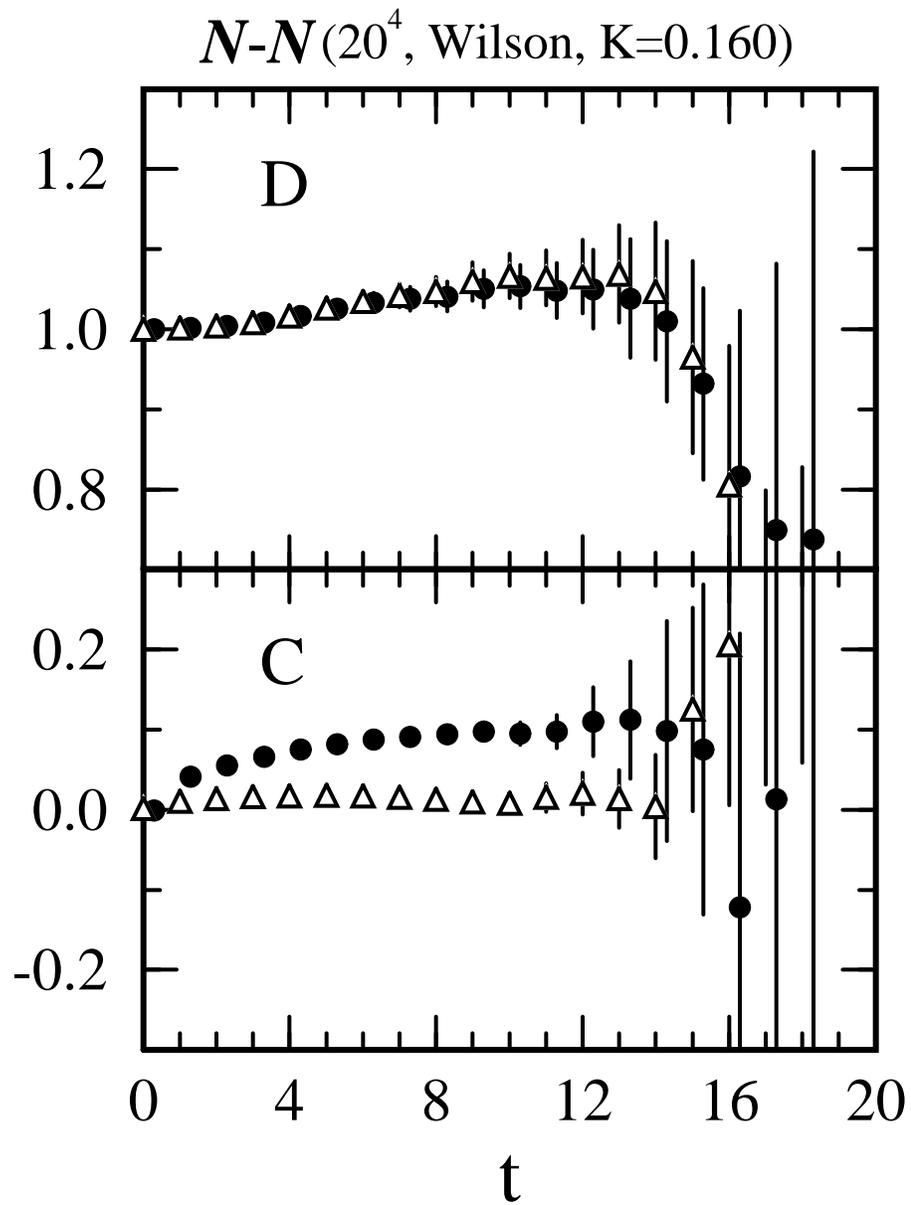

**N-N** ($20^4$, Wilson, K=0.160)

Fig.19

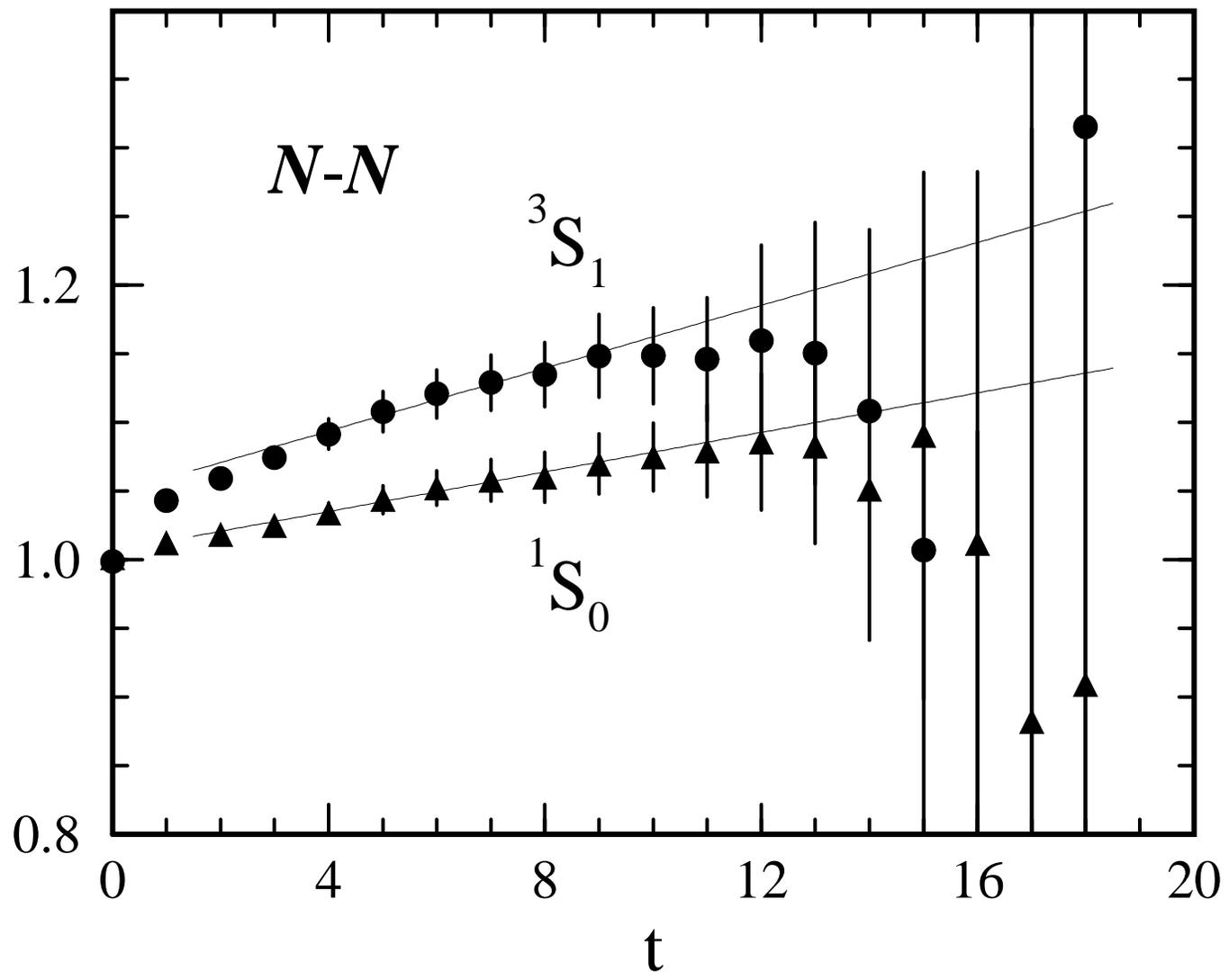

Fig.20

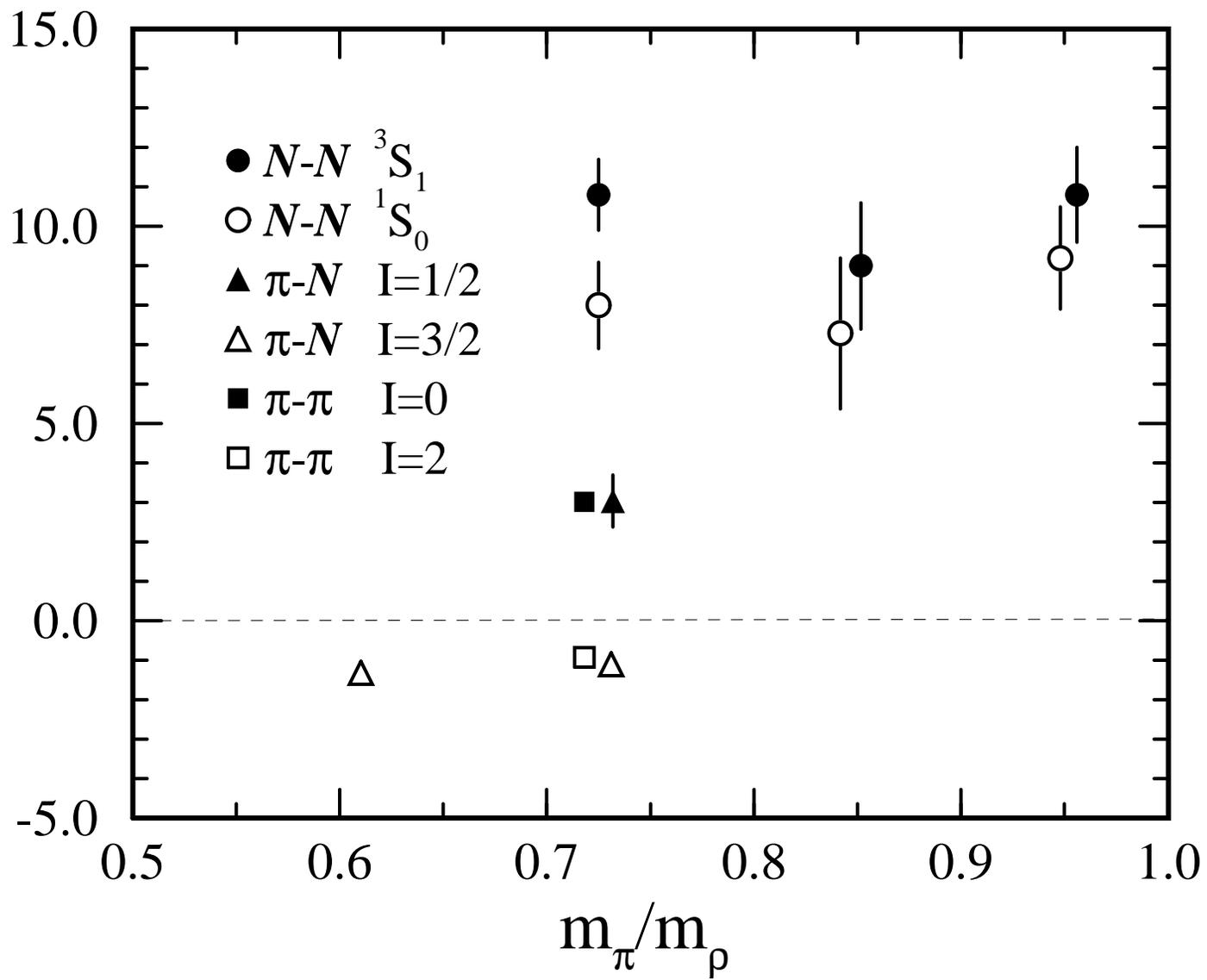

Fig.21

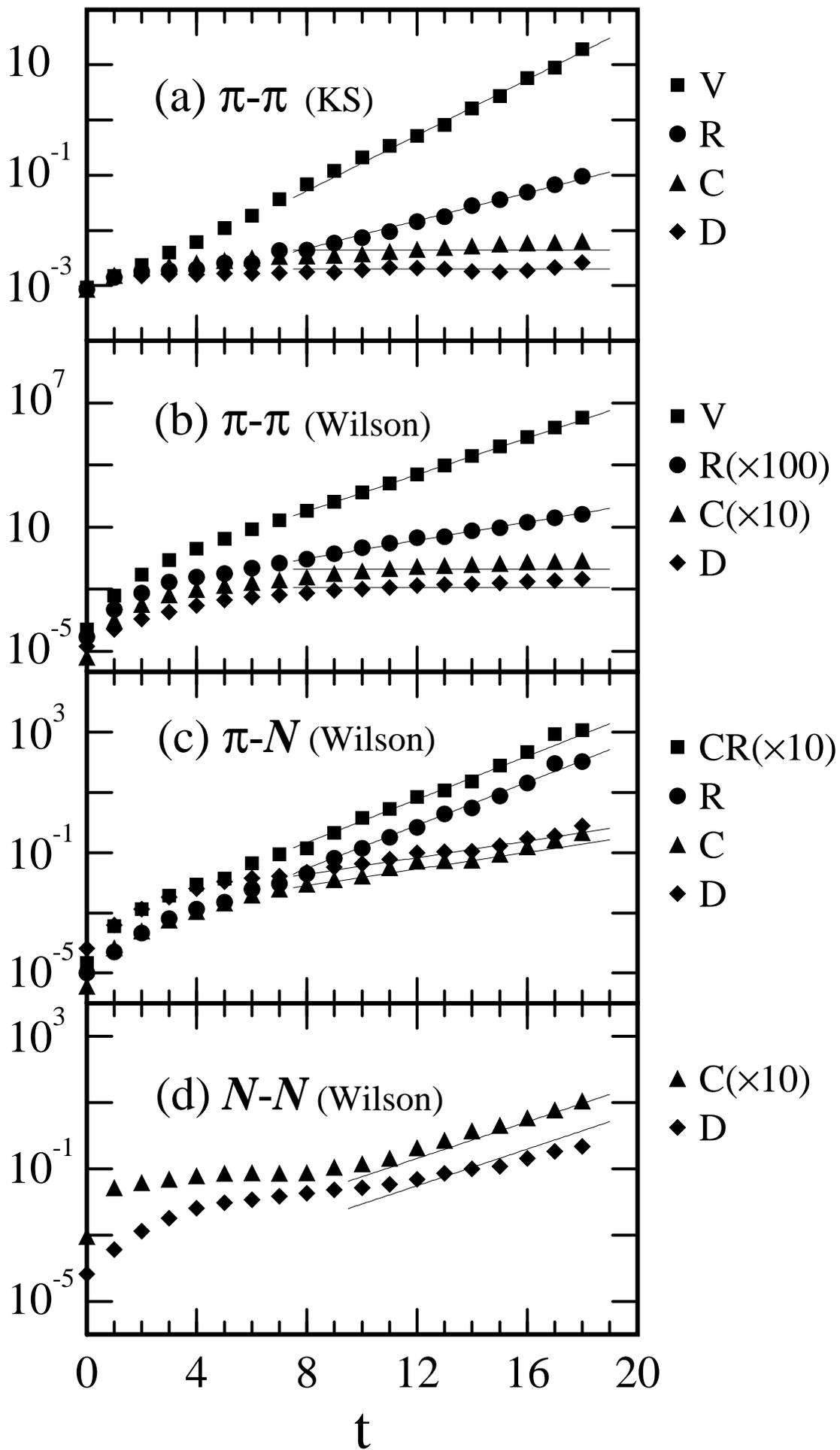

Fig.22

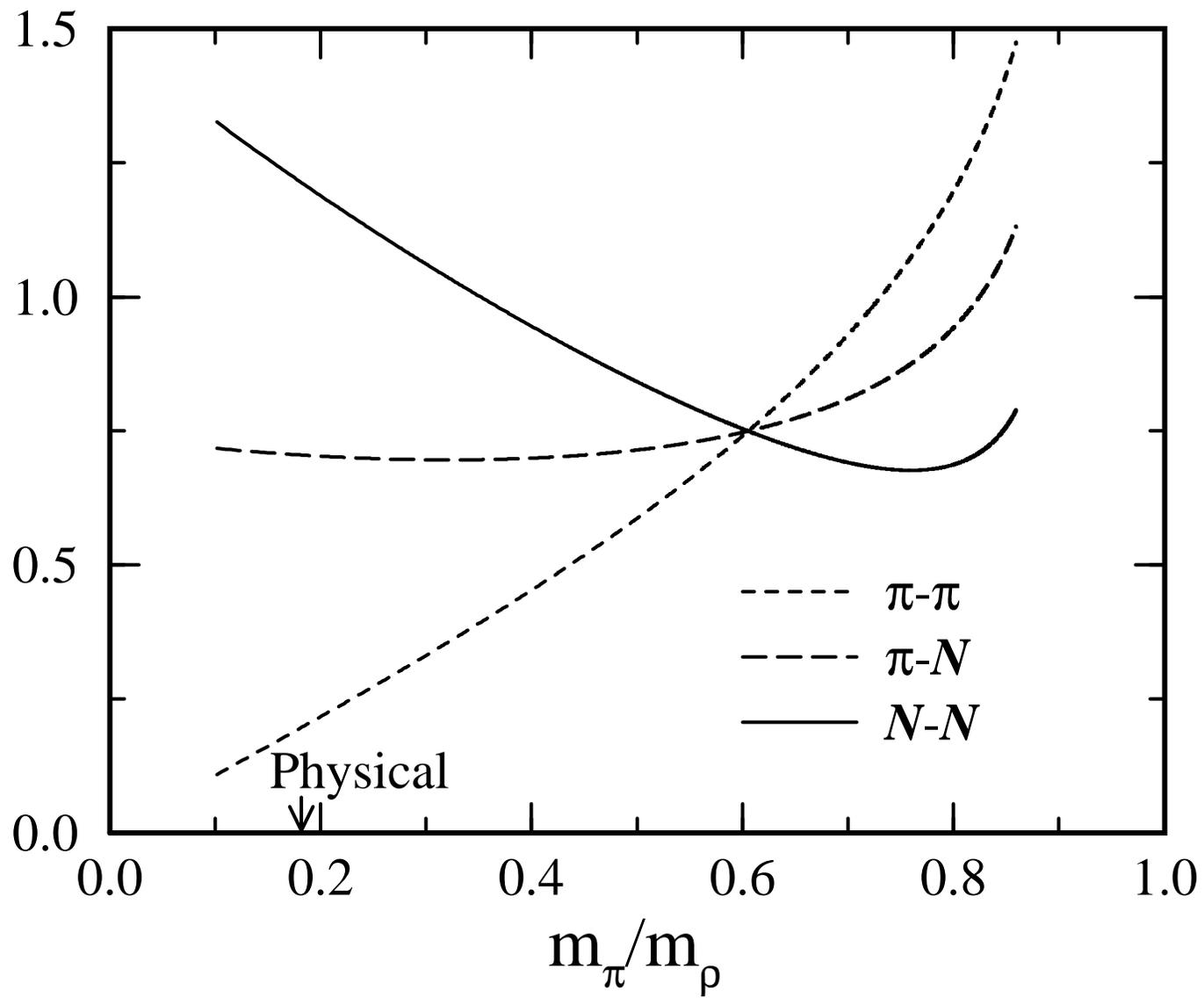

Fig.23

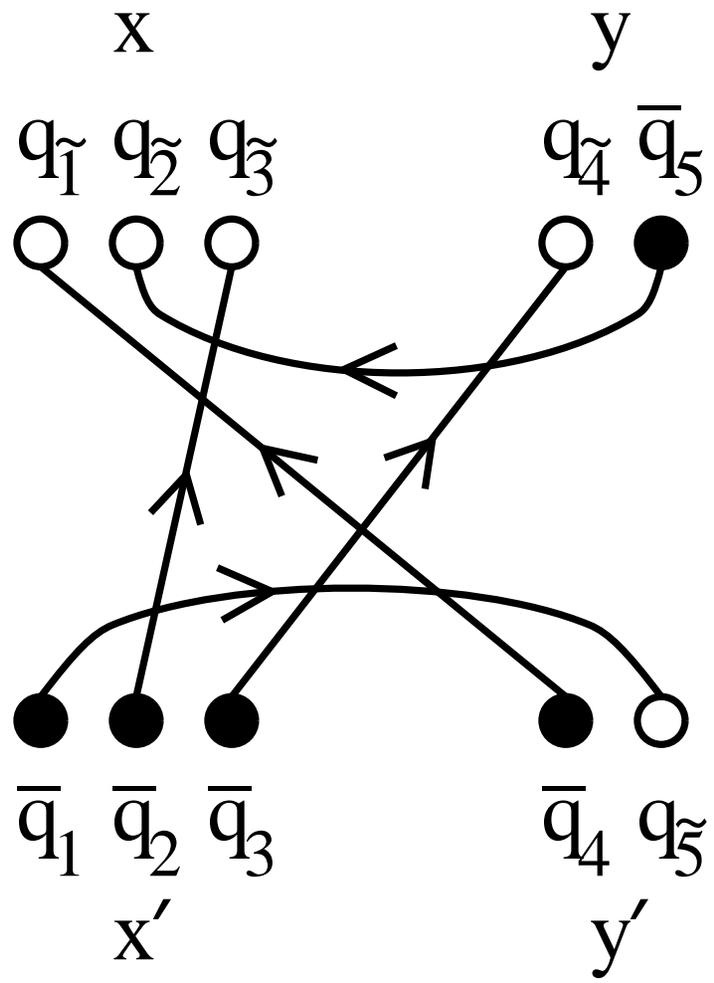

Fig.24

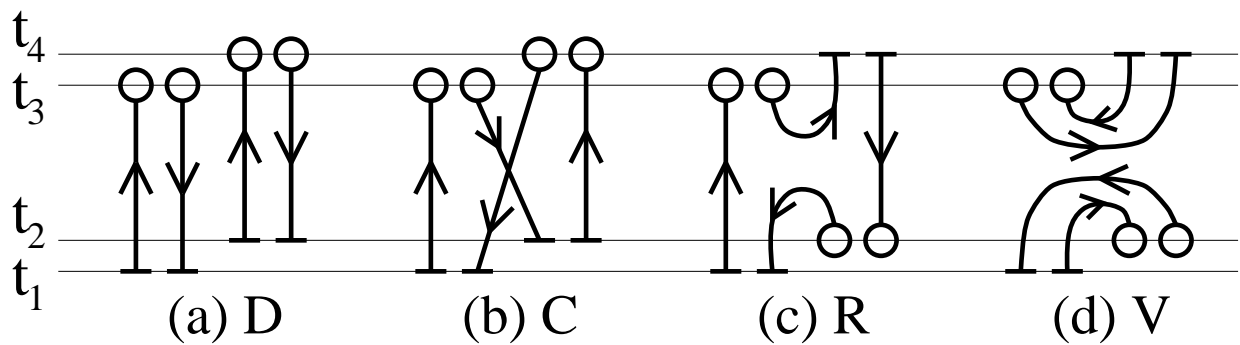

Fig.1

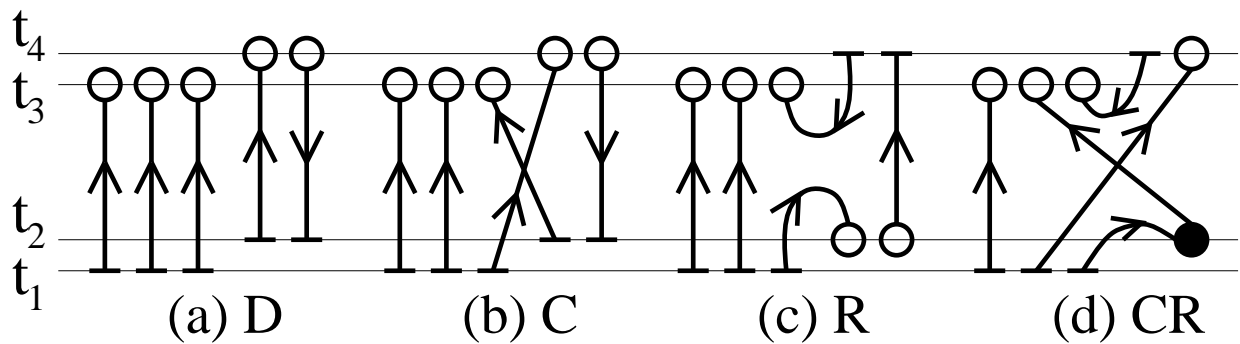

Fig.2

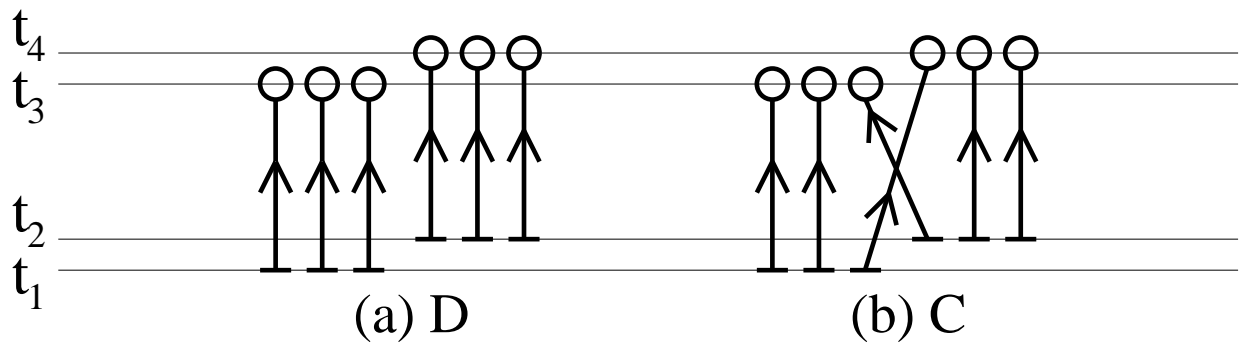

Fig.3